\documentclass[11pt,a4paper]{article}
\pdfoutput=1
\usepackage{jcappub}
\usepackage{amsmath}
\usepackage{graphicx}
\usepackage{epstopdf}
\usepackage{latexsym}
\usepackage{amssymb}

\usepackage{hyperref}
\hypersetup{
    colorlinks=true,
    linkcolor=blue,
    citecolor=blue,
}
\usepackage{bm}
\usepackage{textpos}
\usepackage{caption}
\usepackage{subcaption}
\usepackage{appendix}
\allowdisplaybreaks
\usepackage{wrapfig}
\usepackage{sidecap}
\usepackage{lscape}
\usepackage{array}
\usepackage{tensor}
\usepackage[utf8]{inputenc} 

\begin{document}

\title{Probing the imprints of generalized interacting dark energy on the growth of perturbations}

\author[a]{Jurgen Mifsud}
\author[a]{and Carsten van de Bruck}
\affiliation[a]{Consortium for Fundamental Physics, School of Mathematics and Statistics, University of Sheffield, Hounsfield Road, Sheffield S3 7RH, UK} 
\emailAdd{jmifsud1@sheffield.ac.uk}
\emailAdd{c.vandebruck@sheffield.ac.uk}  

\date{\today}

\abstract{
We extensively study the evolution and distinct signatures of cosmological models, in which dark energy interacts directly with dark matter. We first focus on the imprints of these coupled models on the cosmic microwave background temperature power spectrum, in which we discuss the multipole peak separation together with the integrated Sachs--Wolfe effect. We also address the growth of matter perturbations, and disentangle the interacting dark energy models using the expansion history together with the growth history. We find that a disformal coupling between dark matter and dark energy induces intermediate--scales and time--dependent damped oscillatory features in the matter growth rate function, a unique characteristic of this coupling. Apart from the disformal coupling, we also consider conformally coupled models, together with models which simultaneously make use of both couplings. 
}

\maketitle

\section{Introduction}
A plethora of cosmological observations are in agreement that our Universe is undergoing an era of accelerated expansion, as confirmed first by observations of supernovae at high redshift \cite{Perlmutter:1998np,Riess:1998cb}. The theoretical understanding of this scientific milestone remains one of the most important open challenges in modern cosmology. Assuming the validity of General Relativity at the largest observed scales, the late--time accelerated expansion of the Universe can not be explained by standard matter which satisfies the strong energy condition. Instead, a cosmic fluid with a generous negative value of pressure to energy density ratio is needed to drive this accelerated cosmic expansion. This new energy species is dubbed dark energy. According to the current cosmological observations, dark energy is the major constituent of the present energy budget of the Universe, making up approximately sixty--nine percent \cite{Ade:2015xua} of the overall energy content in the Universe.

In its simplest form, dark energy could be conveniently provided by a non--dynamical positive cosmological constant which peculiarly started to dominate the energy budget of the Universe only recently. This gave rise to the concordance $\Lambda$--cold dark matter ($\Lambda$CDM) cosmological model which is in excellent agreement with all current data. According to cosmological observations, the cosmological constant $\Lambda$ needs to be very small. The corresponding energy density is of the order of a few $\text{meV}^4$, which is very small compared to energy scales expected from particle physics. More general dynamical forms of energy are allowed by data, for instance one can consider the extensively studied quintessence models \cite{Peccei:1987mm,Wetterich:1987fm,Peebles:1987ek}. In these models, the accelerated expansion of the Universe is driven by the dynamics of a scalar field. At late--times, the evolution of the scalar field is primarily dominated by the potential energy rather than its kinetic energy, and the corresponding mass of the field will typically be of the order of $10^{-33}\,\text{eV}$.

The other constituent of the unknown dark sector of the Universe is in the form of cold non--baryonic dark matter, which together with dark energy make up ninety--five percent of the total energy of the Universe. Thus, we are now faced by another challenge of the so called coincidence problem, which deals with the puzzle of having the current energy densities of the dark sector elements with the same order of magnitude. The $\Lambda$CDM model is not able to address this issue, although the tracking nature \cite{Zlatev:1998tr} of quintessence models goes some way towards the resolution of this problem.

Moreover, dark energy and dark matter are usually assumed to be non--interacting and independent components of the dark sector. However, in light of the exotic nature of the dark sector, there is no fundamental reason to suppress or even forbid this direct coupling (see for example Ref. \cite{DAmico:2016jbm} for a quantum field theory formulation). For instance, from solar system \cite{Ip:2015qsa} and laboratory \cite{Brax:2015hma,Brax:2016did} tests, we know that a coupling between the baryonic sector, which amounts to five percent of the total energy budget of the Universe, and dark energy is severely constrained, although this does not follow for the dark matter species. Several consequences, including the variation of the electromagnetic fine--structure constant \cite{vandeBruck:2015rma}, the rotation in the direction of the polarization of light \cite{Carroll:1998zi}, spectral distortions of the cosmic microwave background radiation \cite{vandeBruck:2012vq,Carsten}, and the emission of Cherenkov and bremsstrahlung radiation from charged particles \cite{vandeBruck:2016cnh} have been studied in the literature. Thus, in order to avoid such rigid constraints, we will be considering a coupled quintessence \cite{Wetterich:1994bg} cosmological model in which the standard model particles are uncoupled from the dark sector interaction. Several couplings of this type have been proposed in the literature (see for example Ref. \cite{Bolotin:2013jpa} for a review) and their cosmological consequences have been thoroughly studied \cite{Zimdahl:2001ar,Farrar:2003uw,Amendola:2003wa,Maccio:2003yk,Koivisto:2005nr,Lee:2006za,Copeland:2006wr,Guo:2007zk,Mainini:2007ft,Brookfield:2007au,Baldi:2008ay,Bean:2008ac,Tarrant:2011qe,Pourtsidou:2013nha,Pettorino:2013oxa,Zumalacarregui:2012us,Xia:2013nua,Jack,Koivisto,Sakstein:2014aca,Sakstein:2014isa,Ade:2015rim,Gleyzes:2015pma,Skordis:2015yra,vandeBruck:2016hpz,Mifsudjm:2017}.

In this work, we will be focusing on a coupled quintessence model in which cold dark matter is coupled with the dark energy scalar field via a conformal and a disformal interaction \cite{Jack}, both of which will be explicitly specified in section \ref{sec:Model}. The conformal interaction term is the well--known conformal transformation widely used as a solution--generating technique, and characterises the Jordan--Fierz--Brans--Dicke class of scalar--tensor theories \cite{Faraoni:1998qx}. The disformal coupling term \cite{Bekenstein} brings along intriguing features which distinguishes it from the pure conformal coupling term, such as the distortion of light cones. The disformal coupling term features in the most general four--dimensional second order scalar--tensor theory \cite{Zumalacarregui:2012us,Bettoni:2013diz,Zuma1}, defined by the Horndeski Lagrangian \cite{Horndeski:1974wa}, as well as in non--linear massive gravity theories \cite{Rham1,Rham2}. 

The organization of this paper is as follows. In section \ref{sec:Model}, we introduce our generalized coupled quintessence model and present the dynamical equations governing its background evolution. In sections \ref{sec:CMB_peak} and \ref{sec:ISW_DE_models}, we address the implications of the coupling between the dark sector elements on the cosmic microwave background temperature power spectrum. We first study the temperature power spectrum peak separation using an analytical approach in section \ref{sec:CMB_peak}, in which we compare coupled models with the $\Lambda$CDM model and study any deviations from this model. The contribution of the integrated Sachs--Wolfe effect to the cosmic microwave background temperature power spectrum arising from the coupled quintessence models is discussed in section \ref{sec:ISW_DE_models}. We then turn our attention to the growth history, and present distinctive features of the matter growth rate function in section \ref{sec:growth_history}, together with a discussion on the scale--dependence and time--dependence of the matter growth rate function. The small--scale limit of the perturbation equations is studied in section \ref{sec:Newtonian_limit}, along with analytical solutions of the coupled dark matter density contrast at four non--trivial fixed points. We draw our final remarks and prospective lines of research in section \ref{sec:conclusions}. In Appendix \ref{sec:Perturbations}, we present the coupled quintessence perturbation equations for a generic coupled perfect fluid, covering both the synchronous gauge and the Newtonian gauge.     

\section{The model and its background dynamics}
\label{sec:Model}
\noindent{The Einstein frame description of our scalar--tensor theory is given by the following action:}
\begin{equation}\label{action}
\mathcal{S} = \int d^4 x \sqrt{-g} \left[ \frac{M_{\rm Pl}^2}{2} R - \frac{1}{2} g^{\mu\nu}\partial_\mu \phi\, \partial_\nu \phi - V(\phi) + \mathcal{L}_{SM}\right] + \int d^4 x \sqrt{-\tilde{g}} \mathcal{\tilde{L}}_{DM}\left(\tilde g_{\mu\nu}, \psi\right),
\end{equation}
where $M_\text{Pl}^{-2}\equiv 8\pi G$ such that $M_\text{Pl}=2.4\times 10^{18}$ GeV is the reduced Planck mass. Dark energy (DE) is described by a quintessence scalar field $\phi$, with a potential $V(\phi)$. The uncoupled standard model (SM) particles are described by the Lagrangian $\mathcal{L}_{SM}$, which includes a relativistic sector $(r)$, and a baryon sector $(b)$. Particle quanta of the dark matter (DM) fields $\psi$, follow the geodesics defined by the metric
\begin{equation}\label{disformal_relation}
\tilde g_{\mu\nu} = C(\phi) g_{\mu\nu} + D(\phi)\, \partial_\mu\phi\, \partial_\nu \phi\;, 
\end{equation}
with $C(\phi),\;D(\phi)$ being the conformal and disformal coupling functions, respectively. Throughout this paper we will not be considering a dependence of these functions on the kinetic term $X = -\frac{1}{2} g^{\mu\nu}\partial_\mu \phi \partial_\nu \phi$, although one can extend our results to this more general case \cite{Karwan:2016cnv}. Moreover, the action presented in Eq. (\ref{action}) describes our model in the Einstein frame, which we define to be the frame in which the gravitational sector has the Einstein--Hilbert form, and SM particles are not interacting directly with the quintessence field. Thus, a coupling between DM and DE is induced from the modification of the gravitational field experienced by the DM particles, by the DE scalar field.

We now present the field equations computed from the variation of the action (\ref{action}) with respect to the metric $g_{\mu\nu}$. In this cosmological model, the Einstein field equations take the usual form 
\begin{equation}\label{EFE}
R_{\mu\nu}\,-\,\frac{1}{2}g_{\mu\nu}R = \kappa^2\left(T^\phi_{\mu\nu} + T^{SM}_{\mu\nu} + T^{DM}_{\mu\nu}\right)\;,
\end{equation}
where $\kappa^2\equiv M_\text{Pl}^{-2}$, and the energy--momentum tensors of the scalar field, SM particles, and DM particles are defined by 
\begin{equation}
\begin{aligned}
T^\phi_{\mu\nu}&=\partial_\mu \phi \partial_\nu \phi\,-\,g_{\mu\nu}\left(\frac{1}{2}g^{\rho\sigma}\partial_\rho\phi\partial_\sigma\phi\;\;+\;\;V(\phi)\right)~,\\
T^{SM}_{\mu\nu}&=-\frac{2}{\sqrt{-g}} \frac{\delta\bigl(\sqrt{-g}\mathcal{L}_{SM}\bigr)}{\delta g^{\mu\nu}}\;,\;
T^{DM}_{\mu\nu}=-\frac{2}{\sqrt{-g}} \frac{\delta\bigl(\sqrt{-\tilde{g}}\tilde{\mathcal{L}}_{DM}\bigr)}{\delta g^{\mu\nu}}\;,
\end{aligned}
\end{equation}
respectively. The non--conservation of the scalar field energy--momentum tensor implies the following relation
\begin{equation}\label{eq:modKG}
\Box\phi=V_{,\phi} - Q\;,
\end{equation}
where $V_{,\phi}\equiv dV/d\phi$, and the coupling function is given by
\begin{equation}\label{coupling}
Q=\frac{C_{,\phi}}{2C}T_{DM}+\frac{D_{,\phi}}{2C}T_{DM}^{\mu\nu}\nabla_\mu\phi\nabla_\nu\phi-\nabla_\mu\left[\frac{D}{C}T^{\mu\nu}_{DM}\nabla_\nu\phi\right]\;,
\end{equation}
with $T_{DM}$ being the trace of $T_{DM}^{\mu\nu}$, which, as a consequence of the Bianchi identities, satisfies a modified conservation equation  
\begin{equation}\label{cons}
\nabla^\mu T^{DM}_{\mu\nu}=Q\nabla_\nu\phi\;.
\end{equation}
Since SM particles are not interacting directly with the quintessence field, their energy--momentum tensor obeys the standard conservation equation
\begin{equation}
\nabla^\mu T^{SM}_{\mu\nu}=0\;.
\end{equation}
We consider all species in this model to be described by a perfect fluid energy--momentum tensor
\begin{equation}\label{em_tensor}
T^{\mu\nu}_{i}=(\rho_{i}+p_{i})\bar{u}^\mu \bar{u}^\nu+p_{i} g^{\mu\nu}\;,
\end{equation}
where the index $i$ runs over all the constituents making up the dark and visible sectors. Moreover, we denote the zeroth--order four--velocity of the fluid by $\bar{u}^\mu$, and the Einstein frame SM and DM fluid's energy density and pressure by $\rho_{i}$ and $p_{i}$, respectively. 

We now consider the background evolution of our model in a standard flat Friedmann-Robertson-Walker (FRW) metric, defined by the line element
\begin{equation}\label{FRW}
ds^2 = g_{\mu\nu}dx^{\mu} dx^{\nu} = a^2(\tau)\left[-d\tau^2 + \delta_{ij} dx^i dx^j\right]\;,
\end{equation} 
where $a(\tau)$ is the cosmological scale factor with conformal time $\tau$. In this setting, the modified Klein--Gordon equation, given by Eq. \eqref{eq:modKG}, simplifies to
\begin{equation}\label{KG-equation}
\phi^{\prime\prime} + 2 \mathcal{H} \phi^{\prime} + a^2 V_{,\phi} = a^2 Q\;,
\end{equation}
the fluid conservation equations reduce to
\begin{eqnarray}
\rho_r^\prime + 4\mathcal{H}\rho_r &=& 0\;, \\
\rho_b^\prime + 3\mathcal{H}\rho_b &=& 0\;, \\
\rho_c^\prime + 3\mathcal{H}\rho_c &=& -Q\phi^{\prime}\;,\label{conservation_matter}
\end{eqnarray}
and the Friedmann equations take their usual form
\begin{eqnarray}
\mathcal{H}^2 &=& \frac{\kappa^2}{3}a^2\left(\rho_\phi + \rho_b + \rho_r + \rho_c \right)\;,\label{Friedmann} \\
\mathcal{H}^\prime &=& -\frac{\kappa^2}{6} a^2 \left(\rho_\phi + 3p_{\phi} + \rho_b + 2\rho_r + \rho_c\right)\;.
\end{eqnarray}
We denote coupled DM by a subscript $c$, a conformal time derivative by a prime, and the conformal Hubble parameter by $\mathcal{H}=a^\prime/a$. The scalar field's energy density and pressure have the usual forms of $\rho_\phi={\phi^\prime}^2/\left(2a^2\right)+V(\phi)$ and $p_\phi=\rho_\phi-2V(\phi)$, respectively. The coupling function for a generic coupled perfect fluid with an equation of state $w_c$, as defined by Eq. (\ref{coupling}), simplifies to \cite{vandeBruck:2012vq}
\begin{equation}\label{Q}
Q=-\frac{a^2C_{,\phi}\left(1-3w_c\right) + D_{,\phi}{\phi^\prime}^2 - 2D\left(\frac{C_{,\phi}}{C}{\phi^\prime}^2 + a^2V_{,\phi} + 3\mathcal{H}\left(1+w_c\right)\phi^\prime\right)}{2\left[a^2C + D\left(a^2\rho_c - {\phi^\prime}^2\right)\right]} \rho_c \;.
\end{equation}
This simplifies considerably in the pure conformal case to
\begin{equation}
Q^{(c)}=-\frac{1}{2}\left(\ln C\right)_{,\phi}\left(1-3w_c\right)\rho_c\;,
\end{equation}
in which the coupling function becomes proportional to the energy density of the coupled matter component. 

To be concrete, in this paper the functional form of the couplings and scalar field potential are chosen to be as follows
\begin{equation}\label{coupling_choice}
C(\phi)=e^{2\alpha\kappa\phi}\;,\;\;\;\;D(\phi)=D_M^4 e^{2\beta\kappa\phi}\;,\;\;\;\;V(\phi)=V_0^4 e^{-\lambda\kappa\phi}\;,
\end{equation}
where $\alpha,\,D_M,\,\beta,\,V_0,$ and $\lambda$ are constants.

\begin{figure}[t!]
\centering
  \includegraphics[width=0.49\textwidth]{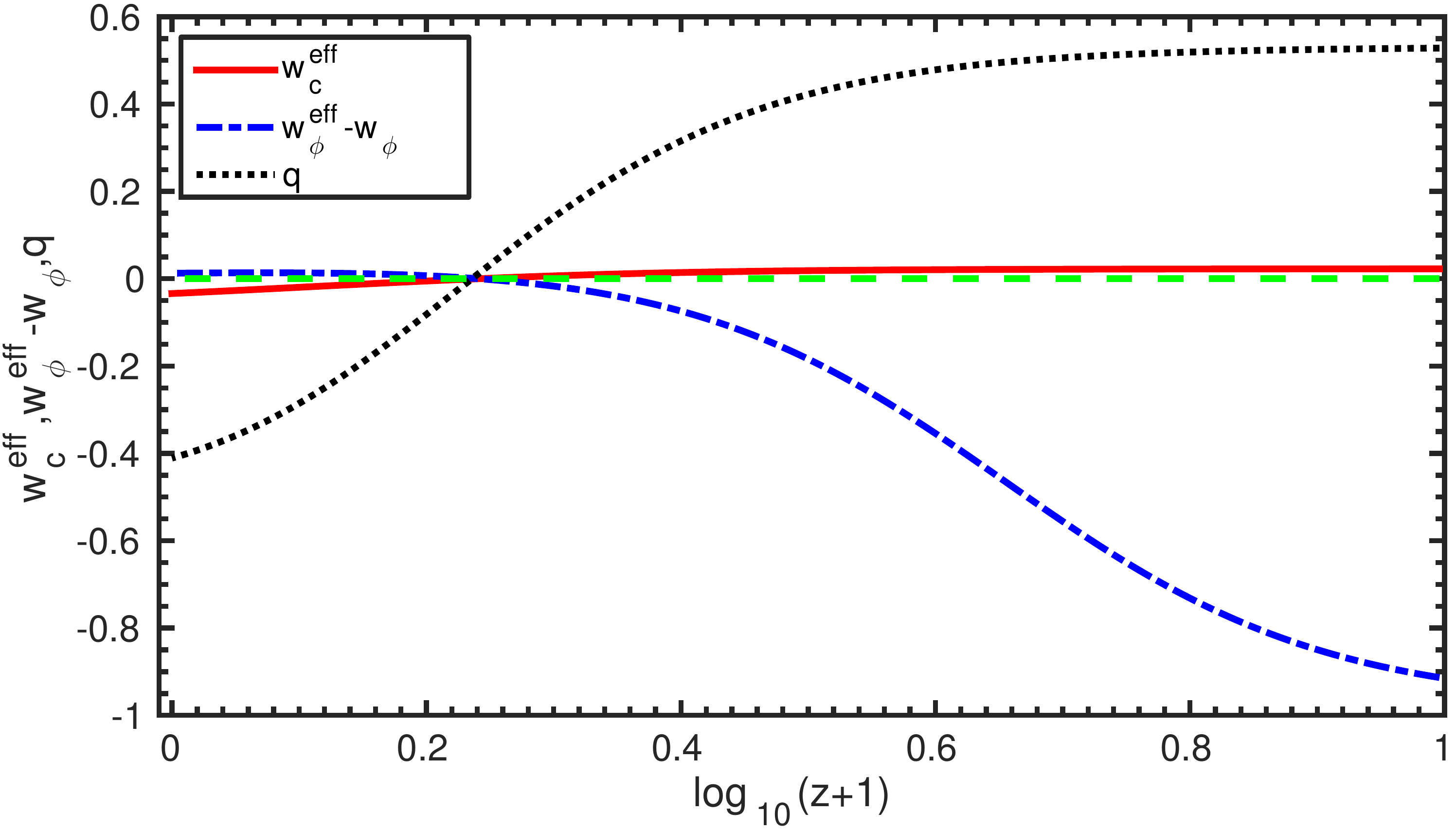}
  \includegraphics[width=0.49\textwidth]{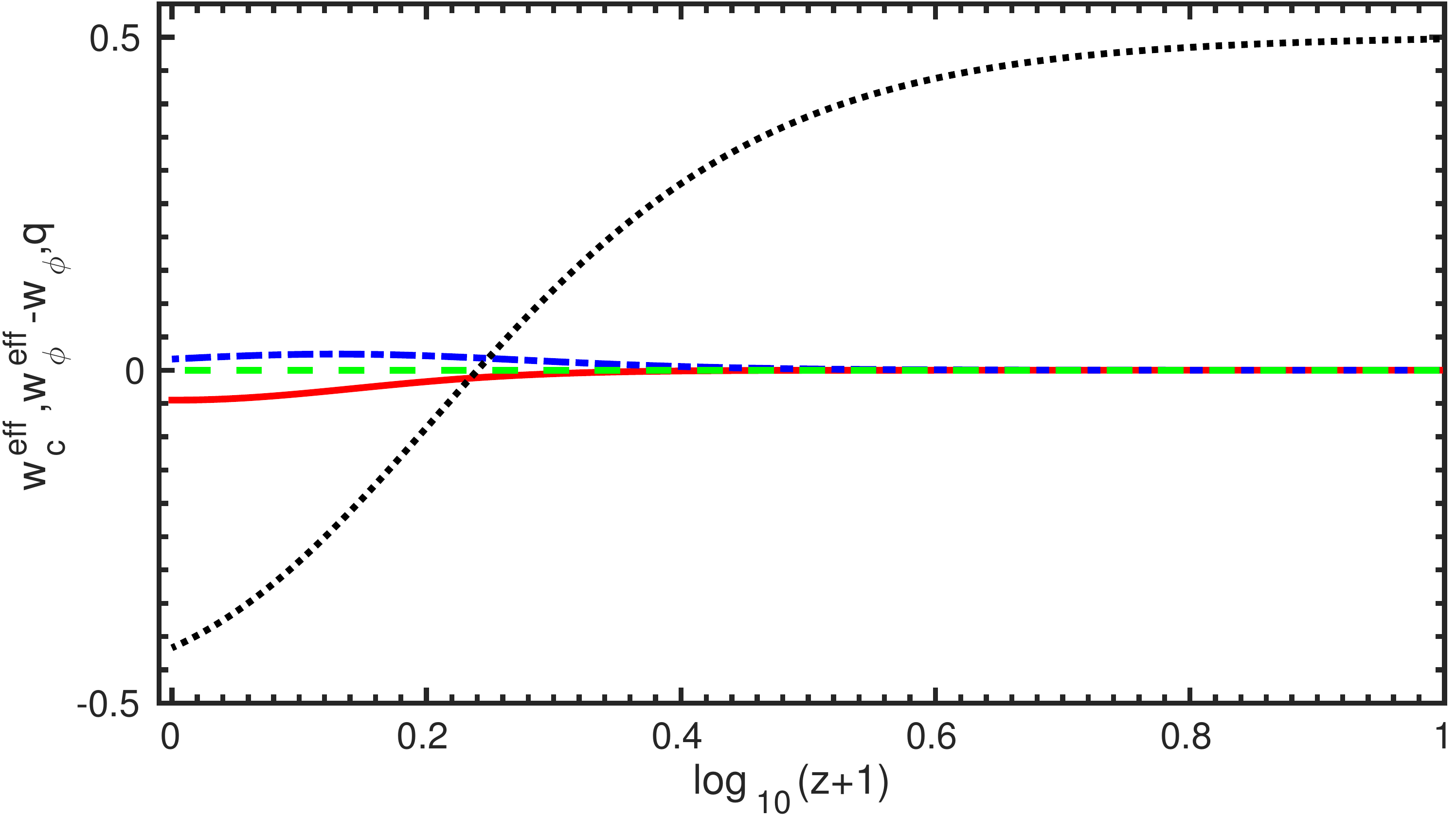}
  \includegraphics[width=0.49\textwidth]{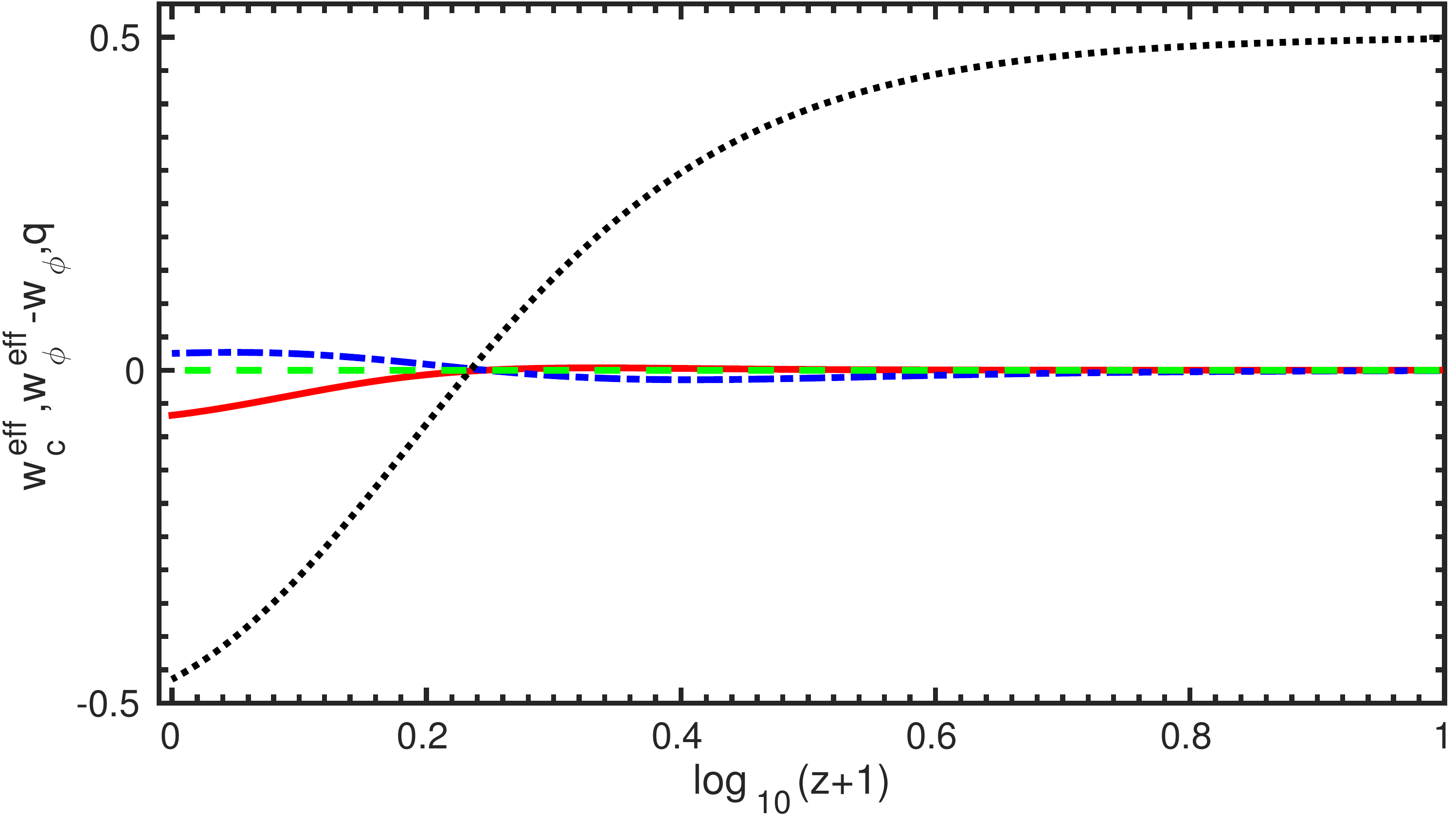}
\caption{These figures show the redshift evolution of $w^{\text{eff}}_c$, $w^{\text{eff}}_{\phi}-w_{\phi}$, and the deceleration parameter $q$, as defined in section \ref{sec:Model}. The couplings and scalar field potential are defined in Eq. (\ref{coupling_choice}). For the conformal case we set $\alpha=0.2$ (top left), for the disformal case we choose $\beta=0$, and $D_M=0.43\,\text{meV}^{-1}$ (top right), and for the mixed case we use $\alpha=0.2$, $\beta=0$, and $D_M=0.43\,\text{meV}^{-1}$ (bottom). In all cases we set $\lambda=1.15$, and depict the abscissa by a dashed line.} 
\label{fig:couplings}
\end{figure}

We can quantify how the coupled DM dilutes with the expansion by rewriting the conservation equation (\ref{conservation_matter}) in terms of a coupling induced effective equation of state for DM
\begin{equation}
\frac{\rho_c^\prime}{\rho_c}+3\mathcal{H}\left(1+w^{\text{eff}}_c\right)=0\;,\;\;\;w^{\text{eff}}_c=\frac{Q\phi^\prime}{3\mathcal{H}\rho_c}\;.
\end{equation}
Similarly, for the scalar field with a pressure to energy density ratio $w_\phi$, we can derive an effective equation of state
\begin{equation}
w^{\text{eff}}_{\phi}=w_{\phi} - \frac{\rho_c}{\rho_{\phi}}w^{\text{eff}}_c\;.
\end{equation}
Hence, when $w^{\text{eff}}_c>0$, DM dilutes faster than in the standard case of $a^{-3}$, and furthermore $w^{\text{eff}}_{\phi}<w_{\phi}$, enhancing the accelerated expansion of the Universe. Conversely, the opposite mechanism takes place when $w^{\text{eff}}_c<0$, leading to an energy flow from DE to DM. We illustrate the evolution of these effective equations of state together with the deceleration parameter $q(z)=-\mathcal{H}^\prime/\mathcal{H}^2$, for three distinct cases in Fig. \ref{fig:couplings}. As expected, the models under consideration give $q(z)<0$ at late--times, leading to a speeding up of the expansion of the Universe, whereas the models give $q(z)>0$ at an earlier epoch, meaning that the expansion was slowed down in the past. As depicted in Fig. \ref{fig:couplings}, the transition redshift is model dependent, although the differences from one model to another are small and depend on the choice of parameters for each respective model. The conformal coupling strength parameter was exaggerated (since cosmological observations forbid such large values \cite{Xia:2013nua,Ade:2015rim,Mifsudjm:2017}) in order to point out that when one introduces a disformal coupling, the energy transfer attributed to the conformal coupling is significantly suppressed. One might think that this makes the model more consistent with cosmological observations, although at the perturbation level, such a model would be in tension with current observations due to an anomalous enhancement in the growth of matter perturbations. We will discuss the evolution of the perturbations in the sections that follow, together with their effects on cosmological observations.
%

\section{An estimation of the separation of CMB peaks}
\label{sec:CMB_peak}
We here estimate the spacing between the peaks in the cosmic microwave background (CMB) temperature power spectrum using only the background evolution of our interacting DE model. It is convenient to define an effective non--interacting DE perfect fluid with energy density $\rho_{\text{DE,eff}}$, and an effective equation of state $w^{\text{eff}}_{\text{DE}}$ \cite{Khoury}, which satisfies the standard conservation equation
\begin{equation}\label{conservation_eff}
\rho_{\text{DE,eff}}^\prime + 3\mathcal{H}\left(1+w^{\text{eff}}_{\text{DE}}\right)\rho_{\text{DE,eff}}=0\;.
\end{equation}
Moreover, in the Friedmann equation, we shall consider a non--interacting DM component
\begin{equation}\label{Friedmann_eff}
\mathcal{H}^2 = \frac{\kappa^2}{3}a^2\left(\rho_{\text{DE,eff}} + \rho_b + \rho_r + \rho_{c,o}a^{-3} \right)\;,
\end{equation}
where $\rho_{c,o}$ is the DM energy density today. By comparing Eq. (\ref{Friedmann_eff}) with Eq. (\ref{Friedmann}), one can easily observe that the evolution of the interacting DE and DM energy densities is absorbed in $\rho_{\text{DE,eff}}$, which is given by
\begin{equation}\label{rho_eff_def}
\rho_{\text{DE,eff}}=\rho_\phi+\rho_c-\rho_{c,o}a^{-3}\;.
\end{equation}
By taking the conformal time derivative of Eq. (\ref{rho_eff_def}), substituting Eq. (\ref{KG-equation}) and Eq. (\ref{conservation_matter}), and comparing the resultant equation with Eq. (\ref{conservation_eff}), one arrives to an expression for the effective equation of state for this effective DE fluid \cite{Khoury} 
\begin{equation}
w^{\text{eff}}_{\text{DE}}=\frac{p_\phi}{\rho_{\text{DE,eff}}}\;.
\end{equation}
\begin{figure}[t!]
\centering
  \includegraphics[width=0.95\textwidth]{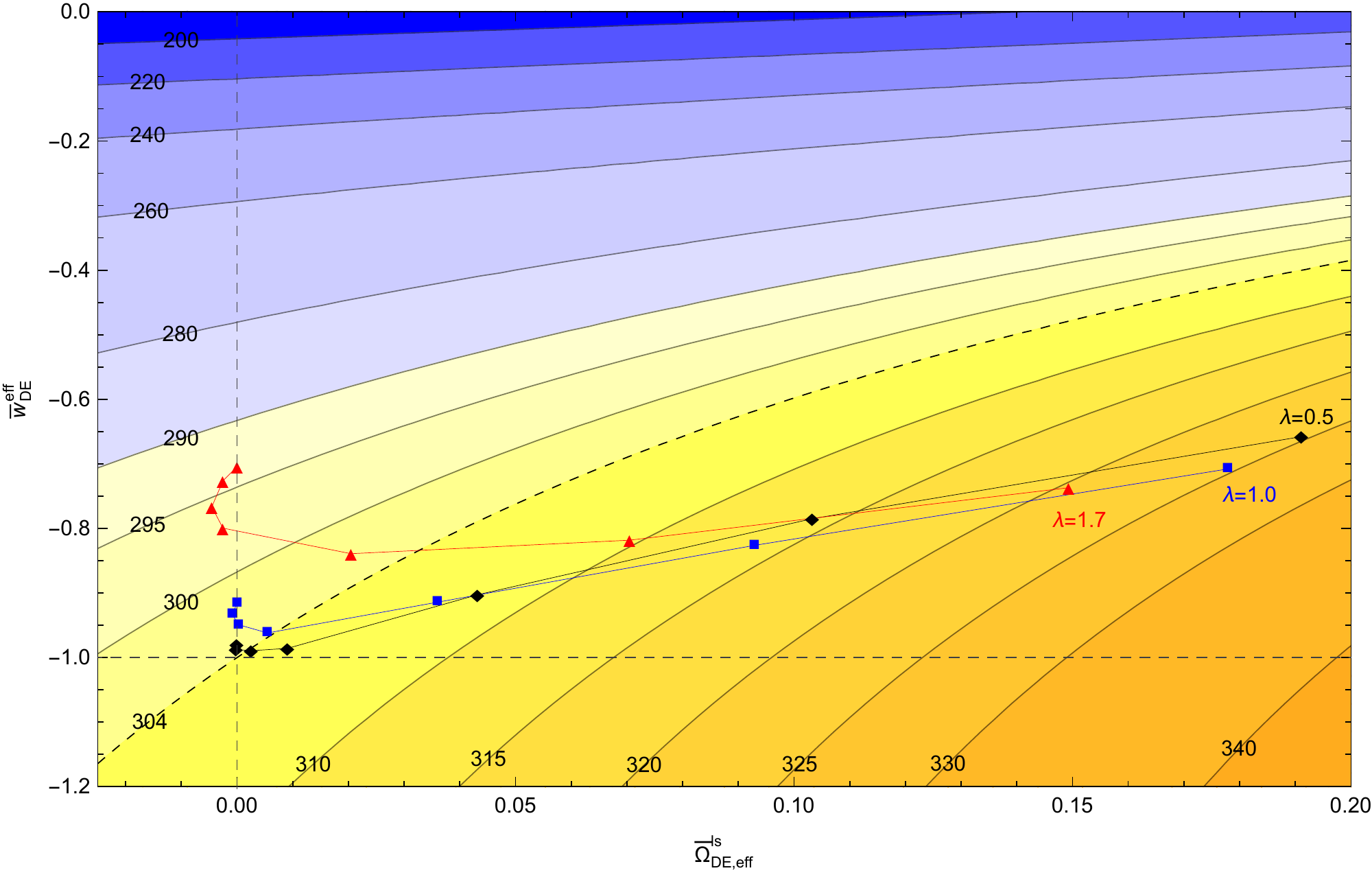}
 \caption{This is a contour plot of the peak separation $\Delta l$, illustrating conformal models with $\lambda=0.5\,(\diamond),\,1.0\,(\square),\,1.7\,(\triangle)$ as a function of $\overline{w}^{\,\text{eff}}_{\text{DE}}$ and $\overline{\Omega}^{ls}_{\text{DE,eff}}$, with $a_{ls}^{-1}=1099.52$ and $\bar{c}_s=0.515$. From right to left, the consecutive points for every choice of $\lambda$ depict conformal models with $\alpha=0.2,\,0.15,\,0.1,\,0.05,\,0.03,\,0.01,\,0$. The $\Lambda$CDM model peak separation is shown by the dashed contour.} 
\label{fig:conformal_contour}
\end{figure}
In order to estimate the spacing between the CMB peaks at different angular momenta $l$, we use the approximation \cite{Hu:1994uz,Doran:2000jt}
\begin{equation}
\Delta l=\pi\frac{\tau_o-\tau_{ls}}{s}=\pi\frac{\tau_o-\tau_{ls}}{\bar{c}_s\tau_{ls}}\;,
\end{equation}
where $\tau_o$ and $\tau_{ls}$ are the conformal time today and at last scattering, respectively. The sound horizon at last scattering is denoted by $s=\bar{c}_s\tau_{ls}$, where the $\tau$-averaged sound speed until last scattering is given by
\begin{equation}
\bar{c}_s=\tau_{ls}^{-1}\int_0^{\tau_{ls}}c_s\,d\tau\;,
\end{equation} 
with the standard sound speed 
\begin{equation}
c_s^{-2}=3+\frac{9}{4}\frac{\rho_b}{\rho_\gamma}\;,
\end{equation}
where $\rho_b/\rho_\gamma$ is the baryon to photon energy density ratio. We now estimate analytically $\tau_o$ and $\tau_{ls}$. For the latter, we consider the interval $0\leq\tau\leq\tau_{ls}$, in which we assume that the fraction of the effective DE $\Omega_{\text{DE,eff}}(\tau)$, does not change rapidly for a considerable period before decoupling. Thus, we can define an effective average 
\begin{equation}
\overline{\Omega}^{ls}_{\text{DE,eff}}=\tau_{ls}^{-1}\int_0^{\tau_{ls}}\Omega_{\text{DE,eff}}\left(\tau\right)d\tau\;,
\end{equation}
with which we can approximate $\Omega_{\text{DE,eff}}(\tau)$ during this period. By solving the Friedmann equation (\ref{Friedmann_eff}), one arrives to an expression for the conformal time at last scattering
\begin{equation}
\tau_{ls}=2H_0^{-1}\left(\frac{1-\overline{\Omega}^{ls}_{\text{DE,eff}}}{\Omega_{b,o}+\Omega_{c,o}}\right)^{\frac{1}{2}}\left[\left(a_{ls}+\frac{\Omega_{r,o}}{\Omega_{b,o}+\Omega_{c,o}}\right)^{\frac{1}{2}} - \left(\frac{\Omega_{r,o}}{\Omega_{b,o}+\Omega_{c,o}}\right)^{\frac{1}{2}} \right]\;,
\end{equation}
where $H_0$ is the Hubble constant, $a_{ls}$ is the cosmological scale factor at last scattering, and $\Omega_{b,o},\,\Omega_{c,o},$ and $\Omega_{r,o}$ are the baryon, DM, and relativistic abundances today.
\begin{figure}[t!]
\centering
  \includegraphics[width=0.95\textwidth]{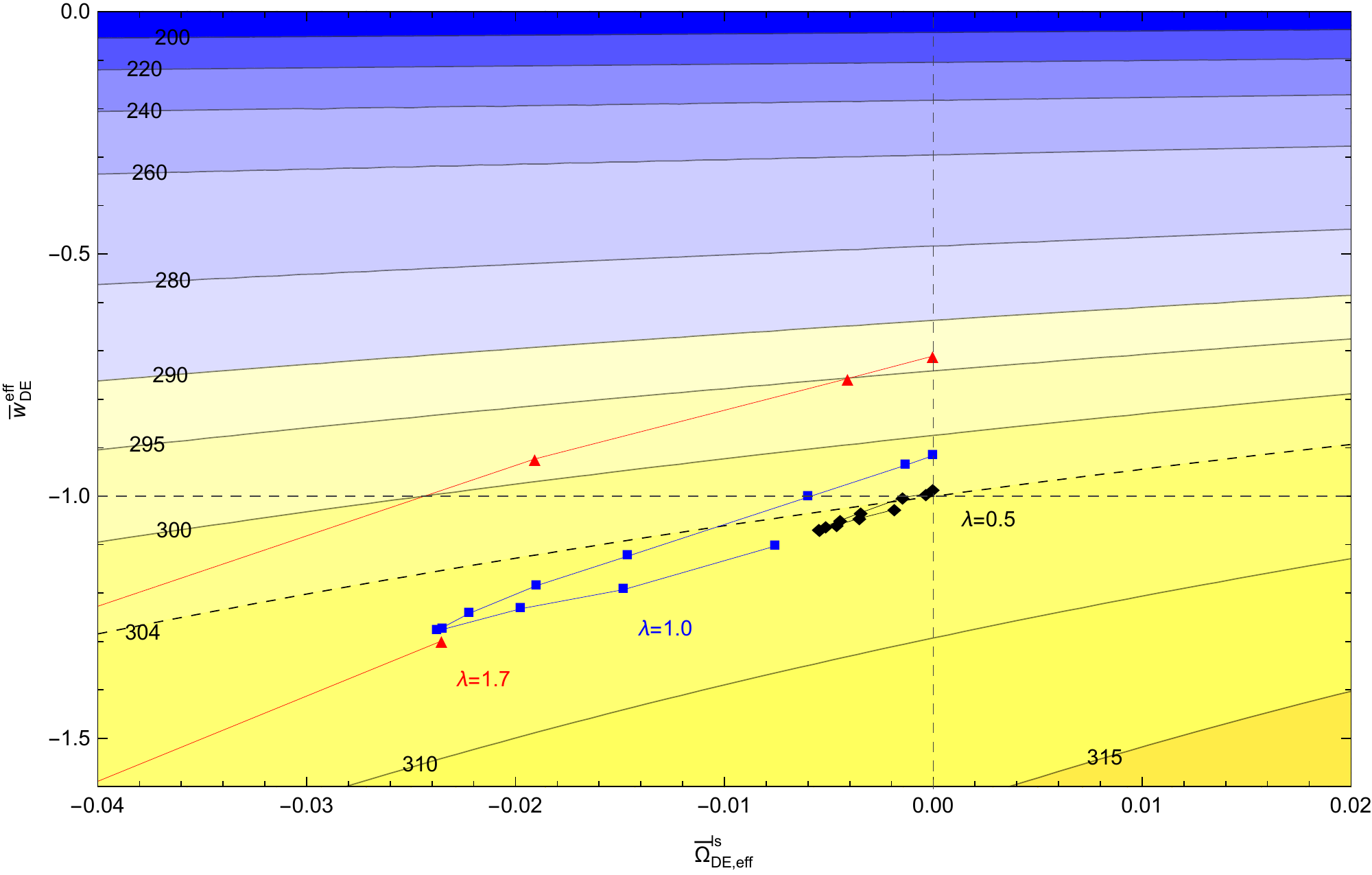}
 \caption{This is a contour plot of the peak separation $\Delta l$, illustrating disformal models with $\lambda=0.5\,(\diamond),\,1.0\,(\square),\,1.7\,(\triangle)$ and $\beta=0$ as a function of $\overline{w}^{\,\text{eff}}_{\text{DE}}$ and $\overline{\Omega}^{ls}_{\text{DE,eff}}$, with $a_{ls}^{-1}=1099.38$ and $\bar{c}_s=0.516$. For each choice of $\lambda$, the consecutive points starting from the $\overline{\Omega}_{\text{DE,eff}}^{ls}$ --axis, depict disformal models with $D_M=0,\,0.2,\,0.3,\,0.4,\,0.45,\,0.5,\,0.55,\,0.6,$ $\,0.7,\,0.8,\,1$ $\text{meV}^{-1}$. The $\Lambda$CDM model peak separation is shown by the dashed contour.} 
\label{fig:disformal_contour}
\end{figure}
We now estimate the conformal time today by considering the interval $0\leq\tau\leq\tau_o$. We define an averaged effective equation of state for the effective DE perfect fluid with energy density $\rho_{\text{DE,eff}}$, as follows
\begin{equation}
\overline{w}^{\,\text{eff}}_{\text{DE}}=\frac{\int_0^{\tau_o}\Omega_{\text{DE,eff}}(\tau)w^{\text{eff}}_{\text{DE}}(\tau)d\tau}{\int_0^{\tau_o}\Omega_{\text{DE,eff}}(\tau)d\tau}\;.
\end{equation}
Thus, for the whole evolution, we estimate the effective equation of state of the effective DE perfect fluid by a constant averaged effective equation of state. From the Friedmann equation (\ref{Friedmann_eff}), one arrives to an expression for the conformal time today
\begin{equation}
\tau_o=2H_0^{-1}\mathcal{F}\;,
\end{equation} 
where
\begin{equation}
\mathcal{F}=\frac{1}{2}\int_0^1\left( \Omega_{\phi,o}a^{1-3\overline{w}^{\,\text{eff}}_{\text{DE}}}+\Omega_{b,o}a+\Omega_{r,o}+\Omega_{c,o}a \right)^{-\frac{1}{2}}da\;,
\end{equation}
with $\Omega_{\phi,o}$ being the DE fraction today. Hence, the CMB peak separation can be estimated by 
\begin{equation}
\Delta l=\pi\bar{c}_s^{\,-1}\left\{\mathcal{F}\left(\frac{\Omega_{b,o}+\Omega_{c,o}}{1-\overline{\Omega}^{ls}_{\text{DE,eff}}}\right)^{\frac{1}{2}}\left[\left(a_{ls}+\frac{\Omega_{r,o}}{\Omega_{b,o}+\Omega_{c,o}}\right)^{\frac{1}{2}} - \left(\frac{\Omega_{r,o}}{\Omega_{b,o}+\Omega_{c,o}}\right)^{\frac{1}{2}}\right]^{-1} -1 \right\}\;.
\end{equation}
We have used the above approach with conformal, disformal, and mixed coupling models, in which we found that this estimation is in very good agreement with the numerical calculations. We compared our estimation with the averaged peak separation over six peaks computed in \texttt{CLASS} \cite{Blas:2011rf} using the full perturbation equations presented in Appendix \ref{sec:Perturbations}. 
\begin{figure}[t!]
\centering
  \includegraphics[width=0.95\textwidth]{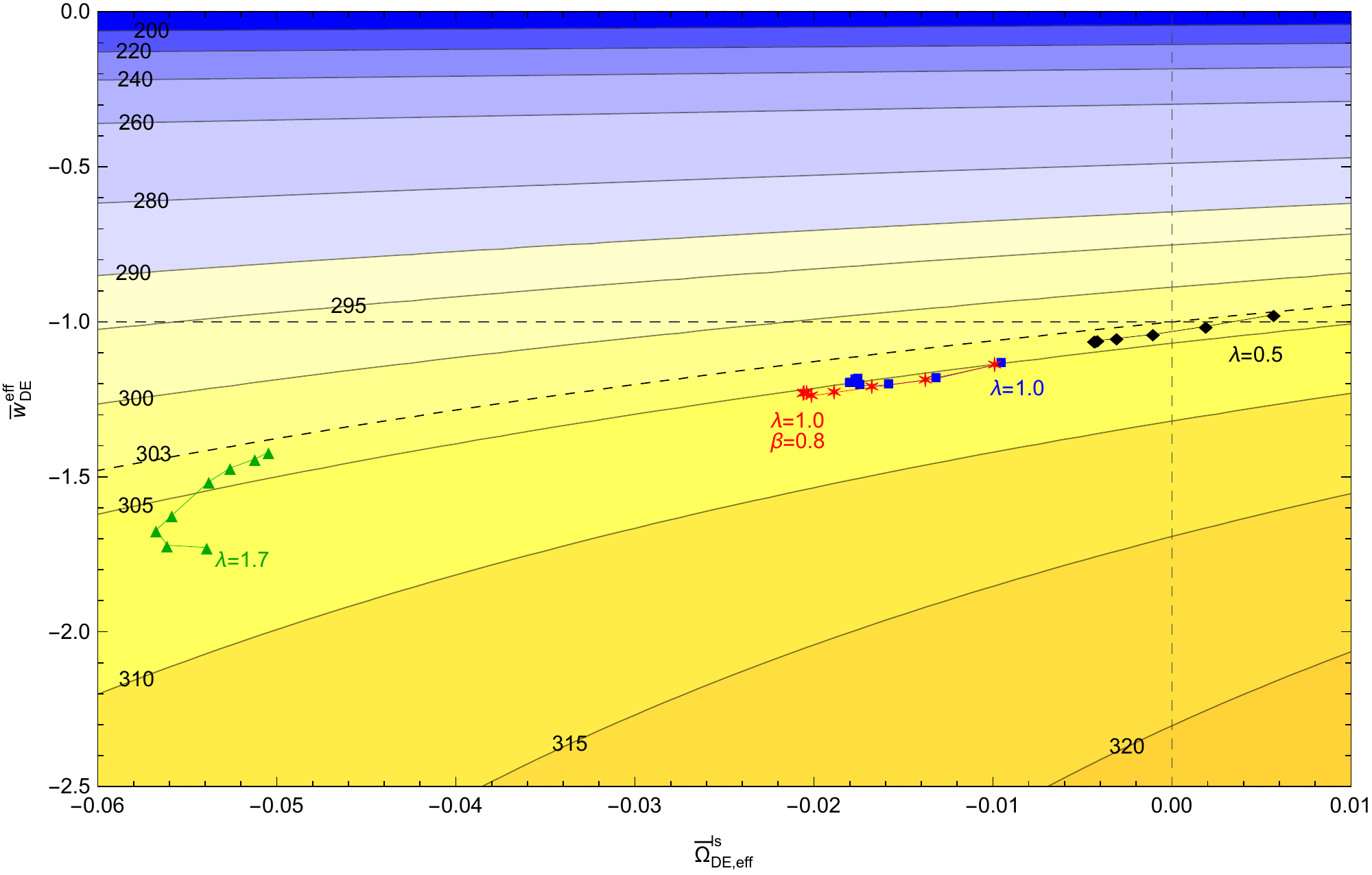}
 \caption{This is a contour plot of the peak separation $\Delta l$, illustrating mixed models with $\beta=0$ and $\lambda=0.5\,(\diamond),\,1.0\,(\square),\,1.7\,(\triangle)$ together with models characterised by $\beta=0.8$ and $\lambda=1.0$ $(\ast)$ as a function of $\overline{w}^{\,\text{eff}}_{\text{DE}}$ and $\overline{\Omega}^{ls}_{\text{DE,eff}}$, with $a_{ls}^{-1}=1096.04$ and $\bar{c}_s=0.515$. From left to right (in a counter--clockwise direction for the points denoted by a $\triangle$), the consecutive points for every choice of $\lambda$ and $\beta$ depict mixed models with $\alpha=0,\,0.01,\,0.03,\,0.05,\,0.1,\,0.15,\,0.2,\,0.25$. For all models, we set $D_M V_0=1$. The $\Lambda$CDM model peak separation is shown by the dashed contour.} 
\label{fig:mixed_contour}
\end{figure}
Indeed, we have checked that when the optimal choice of $a_{ls}$ is chosen for a specific model, the determination of $\Delta l$ is $\lesssim10^{-3}$ percent, and of $\tau_o$ and $\tau_{ls}$ is $\lesssim2-3$ percent. In Figures \ref{fig:conformal_contour}-\ref{fig:mixed_contour} we present contour plots of the CMB peak separation as a function of $\overline{w}^{\,\text{eff}}_{\text{DE}}$ and $\overline{\Omega}^{ls}_{\text{DE,eff}}$ for several parameter choices for the conformal, disformal, and mixed models, respectively. Since every model will have a different value of $a_{ls}$ and $\bar{c}_s$, we have chosen the optimal values of $a_{ls}$ $(\sim 1100^{-1})$ and $\bar{c}_s$ $(\sim 0.52)$ which give the minimal departure from the exact numerical results. The other cosmological parameters have been set to the best fit values reported in Ref. \cite{Ade:2015xua}. In each contour plot, we show the $\Lambda$CDM peak spacing by a dashed contour. 

One can easily notice that the CMB spacing is a robust probe for conformal models, since a larger conformal coupling parameter produces a more pronounced deviation from the $\Lambda$CDM model which currently fits the data very well. Thus, the conformal coupling parameter is easily constrained from the temperature power spectrum of the CMB (see for example Refs. \cite{Pettorino:2013oxa,Xia:2013nua,Ade:2015rim,Mifsudjm:2017}). Indeed, the alteration of the amplitude and the shift of the CMB acoustic peaks to larger multipole moments could be significant as one increases the conformal coupling strength parameter. On the other hand, both the disformal as well as the mixed models are very hard to disentangle from the $\Lambda$CDM model as the CMB peak separation of these models does not deviate significantly from that predicted in the concordance model. Thus, we expect that the parameter space of disformal and mixed models will not be constrained very well from the temperature power spectrum of the CMB alone.  

Another important difference between a purely conformal model and the other interacting models with a disformal coupling, is that in a conformal model the contribution of the effective DE at last scattering $\overline{\Omega}^{ls}_{\text{DE,eff}}$, can be much larger than that in the other models. In conformal models, this non--negligible contribution is coming from the fact that DE starts to contribute even at the time of recombination, thus altering the proportions of DM, baryons, and radiation at decoupling. On the other hand, when a disformal coupling is present, DM, baryons, and radiation follow standard quintessence dynamics for the majority of the cosmic history, and only at very late--times the coupling switches on and modifies the dynamics. These different evolutions of the conformal and the disformal couplings are also behind the fact that conformal models are characterised by a positive $\overline{\Omega}^{ls}_{\text{DE,eff}}$, whereas a disformal coupling tends to be associated with a negative $\overline{\Omega}^{ls}_{\text{DE,eff}}$.  Furthermore, one can assert that conformal models occupy the first quadrant of the $\overline{w}^{\,\text{eff}}_{\text{DE}}-\overline{\Omega}^{ls}_{\text{DE,eff}}$ plane with respect to the origin located at the $\Lambda$CDM model, whereas disformal and mixed models are situated in the third quadrant of the same plane, with a slight overlap between conformal and disformal models in the second quadrant.
\section{The ISW effect in interacting dark energy models}
\label{sec:ISW_DE_models}
\begin{figure}[t!]
\centering
  \includegraphics[width=0.49\textwidth]{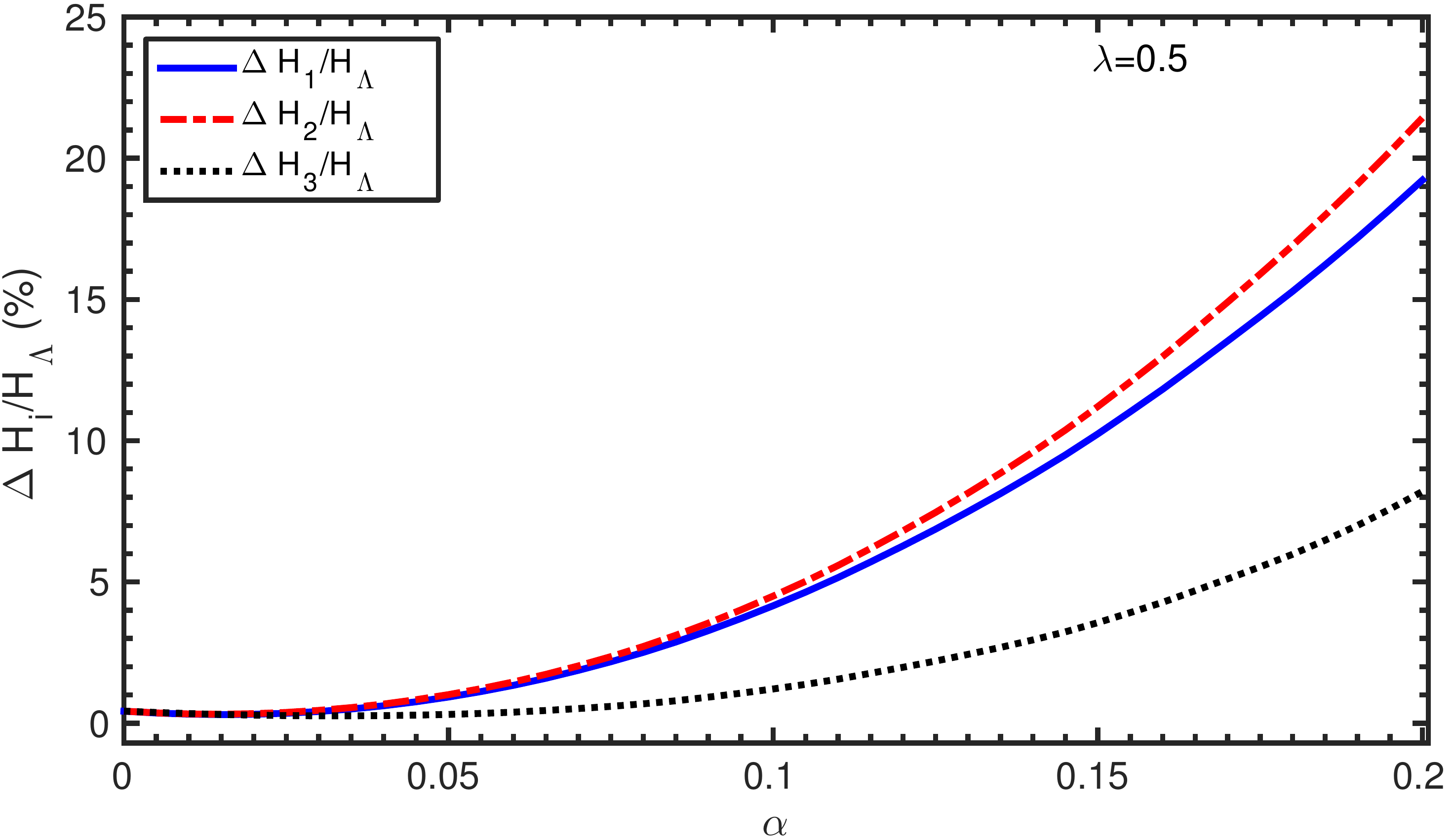}
  \includegraphics[width=0.49\textwidth]{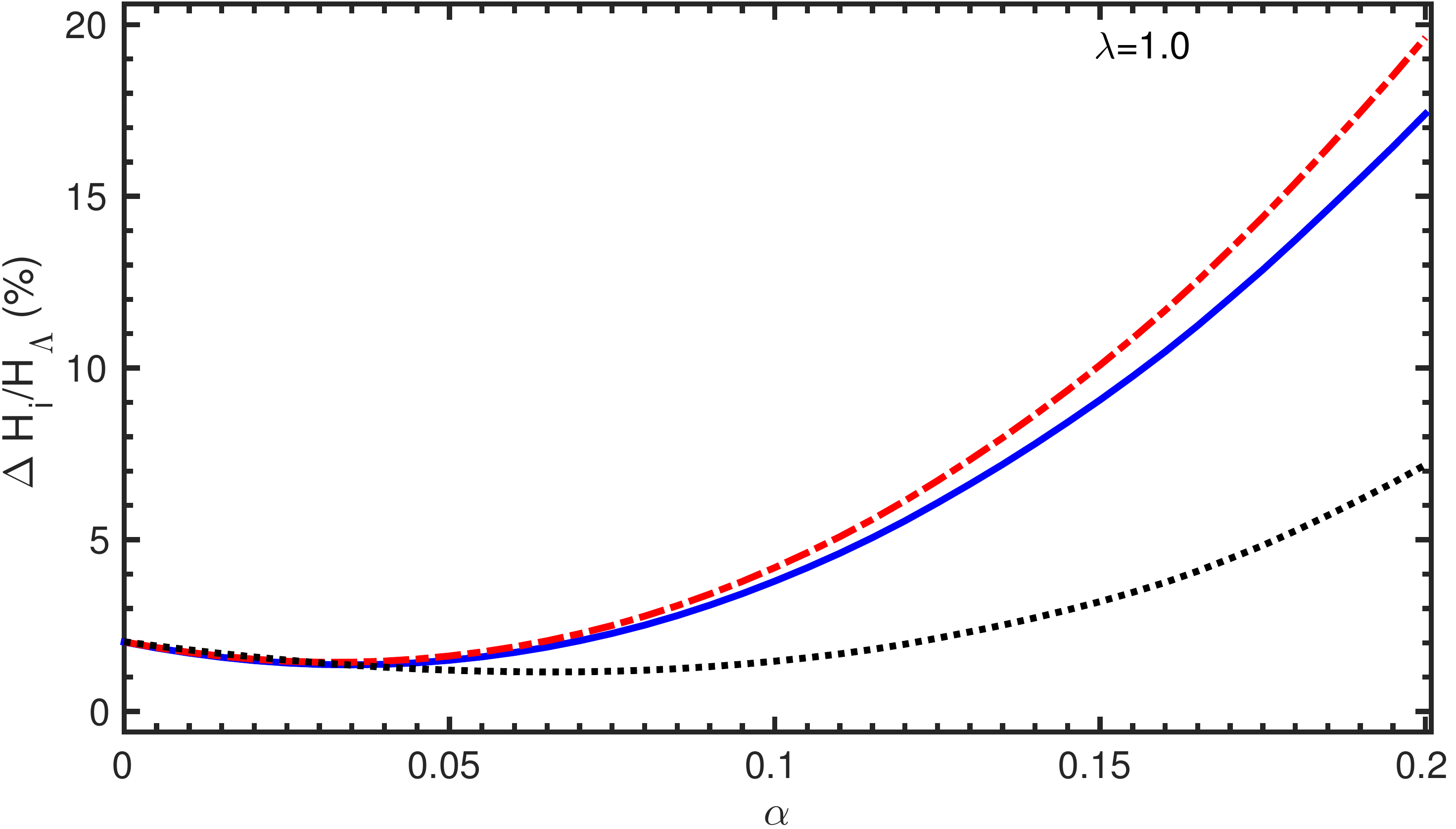} 
 \caption{These figures show the relative difference of $H_1$, $H_2$, and $H_3$ to the $\Lambda$CDM model for conformally coupled models with coupling and potential functions as defined in Eq. (\ref{coupling_choice}). The slope of the potential has been set to $\lambda=0.5$ (left) and to $\lambda=1.0$ (right).} 
\label{fig:Delta_conf}
\end{figure}
We here consider the imprint of interacting DE models on the integrated Sachs--Wolfe (ISW) effect \cite{Sachs:1967er} in the CMB temperature power spectrum which boosts the power at low multipoles. The ISW effect gives a non--zero contribution to the CMB fluctuations whenever the large--scale gravitational potential is time evolving. Thus, this secondary source of CMB anisotropy will not contribute during the matter dominated era, although it will be present after CMB decoupling, and at the very recent times when the expansion of the Universe starts to be dominated by DE. In order to distinguish these interacting DE models from the concordance model, we consider the height of the first three acoustic peaks of the CMB temperature power spectrum relative to the power at $l=10$ by
\begin{equation}
H_i=\left(\frac{\Delta T_{l_i}}{\Delta T_{10}}\right)^2\;,
\end{equation}  
with $i=\{1,2,3\}$, and $\left(\Delta T_{l_i}\right)^2=l_i\left(l_i+1\right)C_{l_i}/2\pi$, where $C_{l_i}$ is the power spectrum of the multipole moments of the temperature field at peak position $l_i$ \cite{Hu:2000ti}. We compare several interacting DE models with the $\Lambda$CDM model with identical Hubble constant, spectral index, baryon density, and DM fraction, by the relative difference of $H_i$ to the $\Lambda$CDM model. We denote this difference by $\Delta H_i/H_\Lambda$, in which we first determine the parameters $H_i$ in the interacting DE model from the CMB spectra, and compare them with those of the $\Lambda$CDM model. 

\begin{figure}[t!]
\centering
  \includegraphics[width=0.49\textwidth]{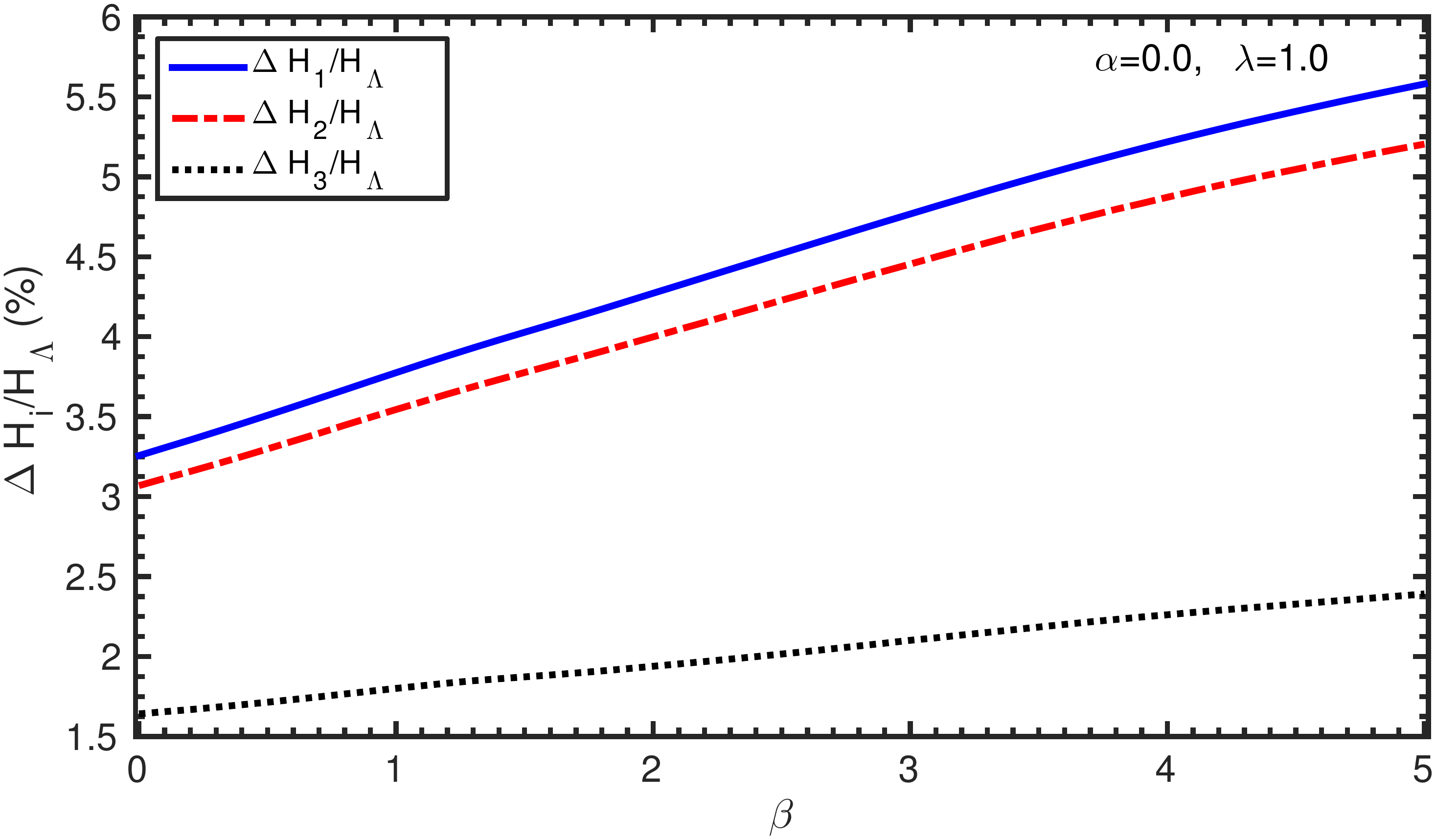}
  \includegraphics[width=0.49\textwidth]{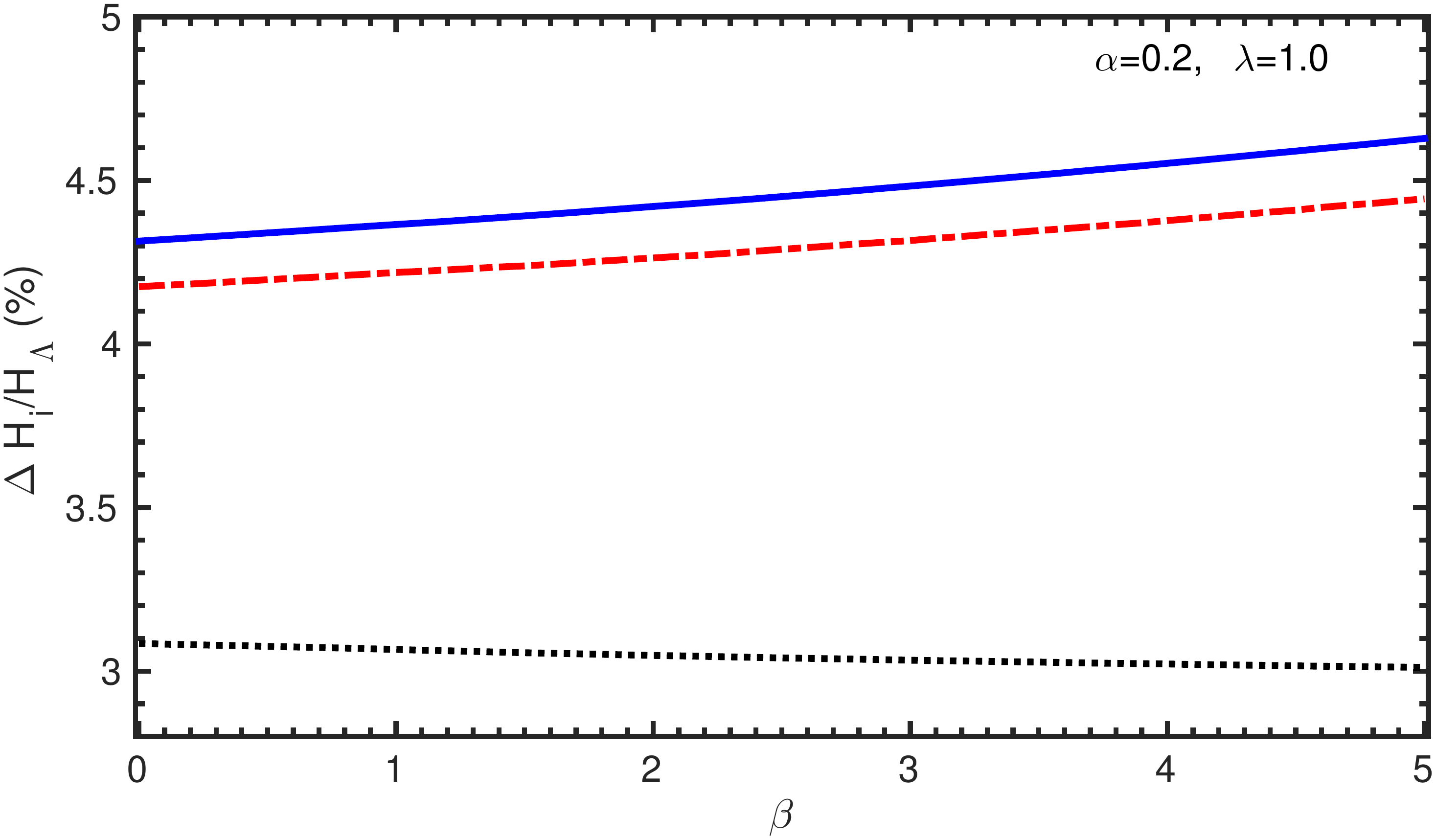} 
 \caption{These figures show the relative difference of $H_1$, $H_2$, and $H_3$ to the $\Lambda$CDM model for disformally coupled models (left) and mixed coupled models (right) with coupling and potential functions as defined in Eq. (\ref{coupling_choice}). For the disformal model (left) we set $\alpha=0.0$ and $\lambda=1.0$, and for the mixed model (right) we use $\alpha=0.2$ and $\lambda=1.0$. In both cases we use the relation $D_M V_0=1$.}  
\label{fig:Delta_mixed}
\end{figure}

We illustrate two conformal models in Fig. \ref{fig:Delta_conf}, and a disformal together with a mixed case in Fig. \ref{fig:Delta_mixed}. In order to distinguish these interacting DE models from the concordance model, we need the relative difference of $H_i$ to be comparable with the dominant uncertainty $(\sim30\%)$ \cite{Hu:2001bc} arising from cosmic variance at $l=10$.
Thus, an immediate observation from the examples presented in Fig. \ref{fig:Delta_conf} and in Fig. \ref{fig:Delta_mixed}, is that both a conformal and a disformal coupling in the dark sector of the Universe are hardly distinguishable from the $\Lambda$CDM model, particularly when a disformal coupling is present. For a pure conformal coupling, the relative difference from the $\Lambda$CDM model increases significantly up to $\sim20$ percent as the coupling strength is enhanced, whereas for disformal and mixed couplings the discrepancy to the $\Lambda$ case stays at the order of a few percent even when the disformal coupling strength is increased considerably. Moreover, a conformal coupling together with a disformal coupling tend to decrease the relative difference of $H_i$ when compared with the pure disformal coupling model, as shown in Fig. \ref{fig:Delta_mixed}. Finally, in conformally coupled models we can see that $H_2$ is the best estimator, whereas $H_1$ gives the largest discrepancy from the $\Lambda$CDM model for the disformally and mixed coupled models, identical to what has been reported for standard quintessence in Ref. \cite{Corasaniti:2002bw}. 

As already mentioned, these best estimators of the ISW effect which give rise to the largest discrepancy between an interacting DE model and the concordance model, are still not able to produce a detectable signature due to the cosmic variance uncertainty. One can overcome this difficulty by cross--correlating matter templates constructed from galaxy catalogues with the CMB temperature power spectrum \cite{Crittenden:1995ak,Olivares:2008bx,Xia:2009zzb,Wang:2016lxa}. This additional probe of the interaction between the dark sector elements could potentially provide further constraints on our model parameters, although this is beyond the scope of this paper.
\section{Imprints on the growth history}
\label{sec:growth_history}
\begin{figure}[t!]
\centering
  \includegraphics[width=0.98\textwidth]{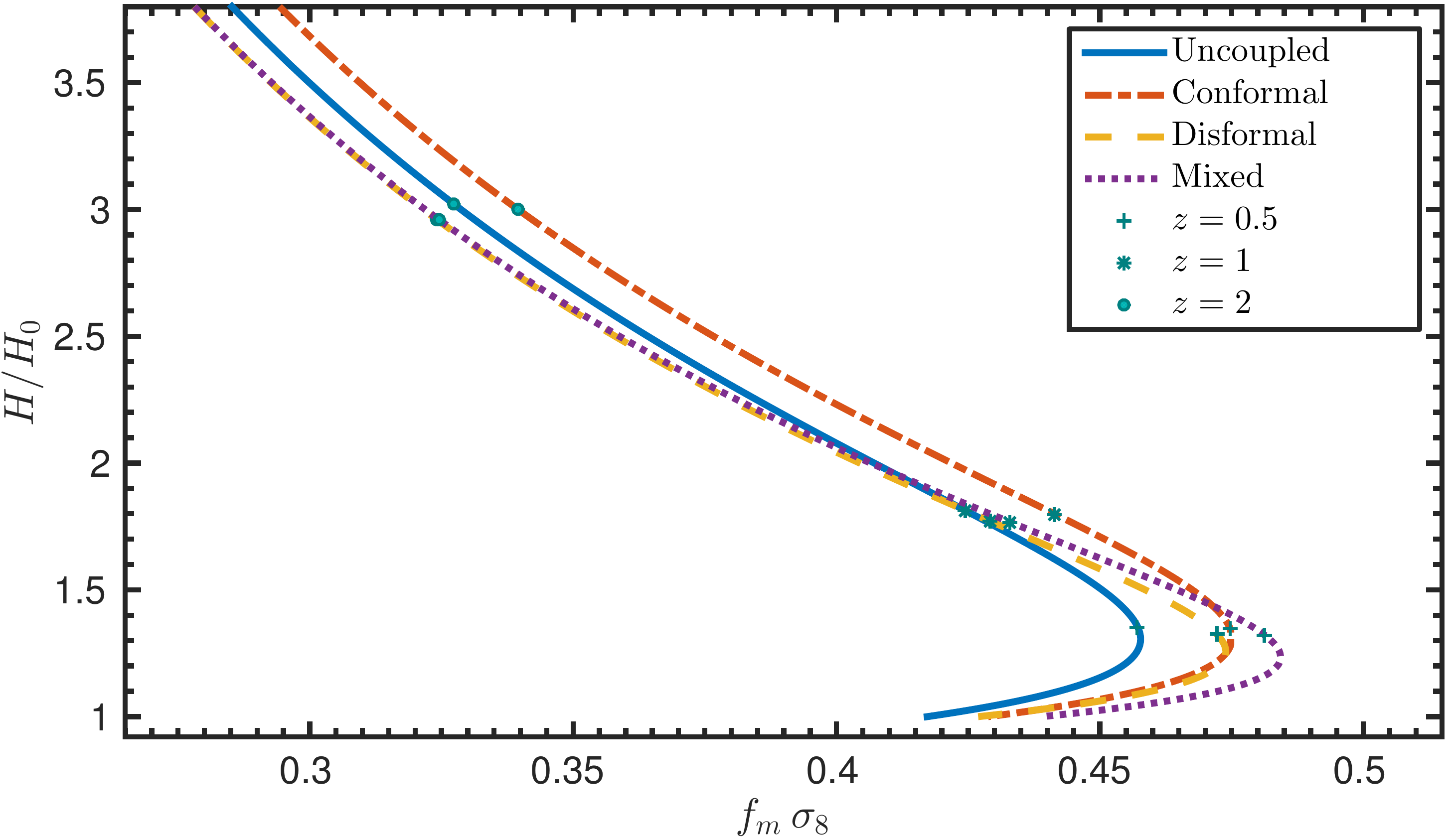}
\caption{This figure shows the expansion history $H/H_0$, against the matter growth history $f_m\sigma_8$, at wave number $k=0.1\,h\,\text{Mpc}^{-1}$. For the conformal model we set $\alpha=0.05$, for the disformal model we choose $D_M=0.43\,\text{meV}^{-1}$ and $\beta=0$, and we use the same parameters in the mixed model. We set $\lambda=1$ in all the models. We depict three specific locations of the redshift along each curve by a $+,\,\ast,\,\circ$ corresponding to $z=0.5,\,1,\,2$, respectively. }  
\label{fig:fs8_HHo}
\end{figure}
In this section we discuss the growth history of these interacting DE models. We consider the matter growth rate function defined by
\begin{equation}
f_m=\frac{d\ln\delta_m}{d\ln a}=\frac{\delta_m^\prime}{\mathcal{H}\delta_m}\;,
\end{equation}
where we define the matter density contrast by
\begin{equation}
\delta_m=\frac{\rho_b\delta_b+\rho_c\delta_c}{\rho_b+\rho_c}\;,
\end{equation}
with $\delta_b,\,\delta_c$ being the baryon and coupled DM density contrasts, respectively. In order to distinguish between the interacting DE models, we consider a useful combination of the product of the matter growth rate function $f_m$, with the root mean square mass fluctuation amplitude in spheres of radius $8\,h^{-1}\text{Mpc}$, $\sigma_8(z)$ \cite{Linder:2016xer}. In Fig. \ref{fig:fs8_HHo}, we plot the expansion history against the growth history, more specifically $H/H_0$ against $f_m\,\sigma_8$, where $H=a^{-1}\mathcal{H}$ and $H_0=100\,h\,\text{km}\,\text{s}^{-1}\,\text{Mpc}^{-1}$. The redshift in Fig. \ref{fig:fs8_HHo} runs along the curves, such that it monotonically decreases from top to bottom. Thus, by locating the same redshift on each curve, one can determine if the expansion rate differs from one model to another. In this figure we locate three different redshifts on each curve, and one can easily observe that at any given redshift these models give a different value of $H/H_0$, although the difference is small.
\begin{figure}[t!]
\centering
  \includegraphics[width=0.48\textwidth]{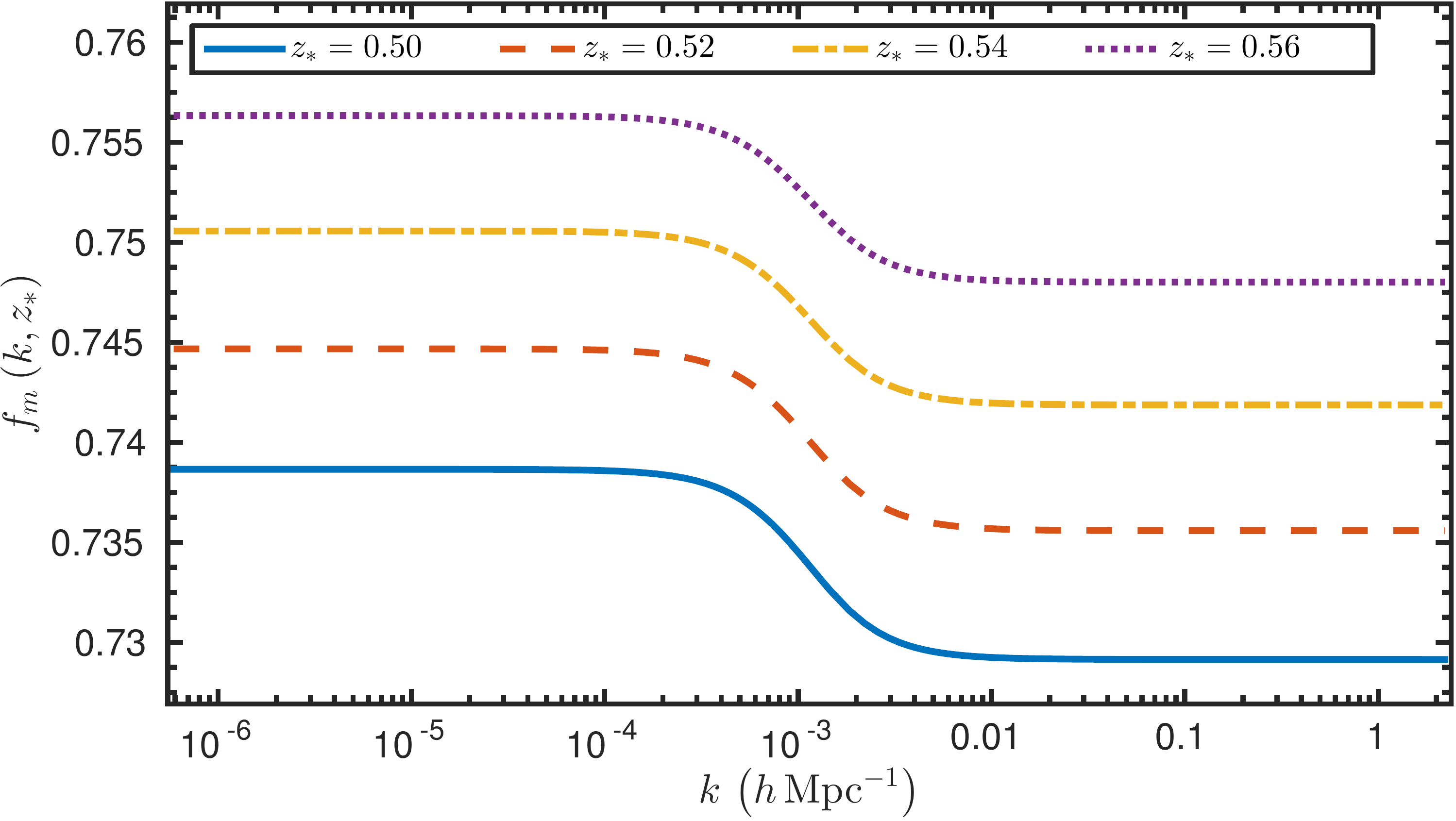}
  \includegraphics[width=0.48\textwidth]{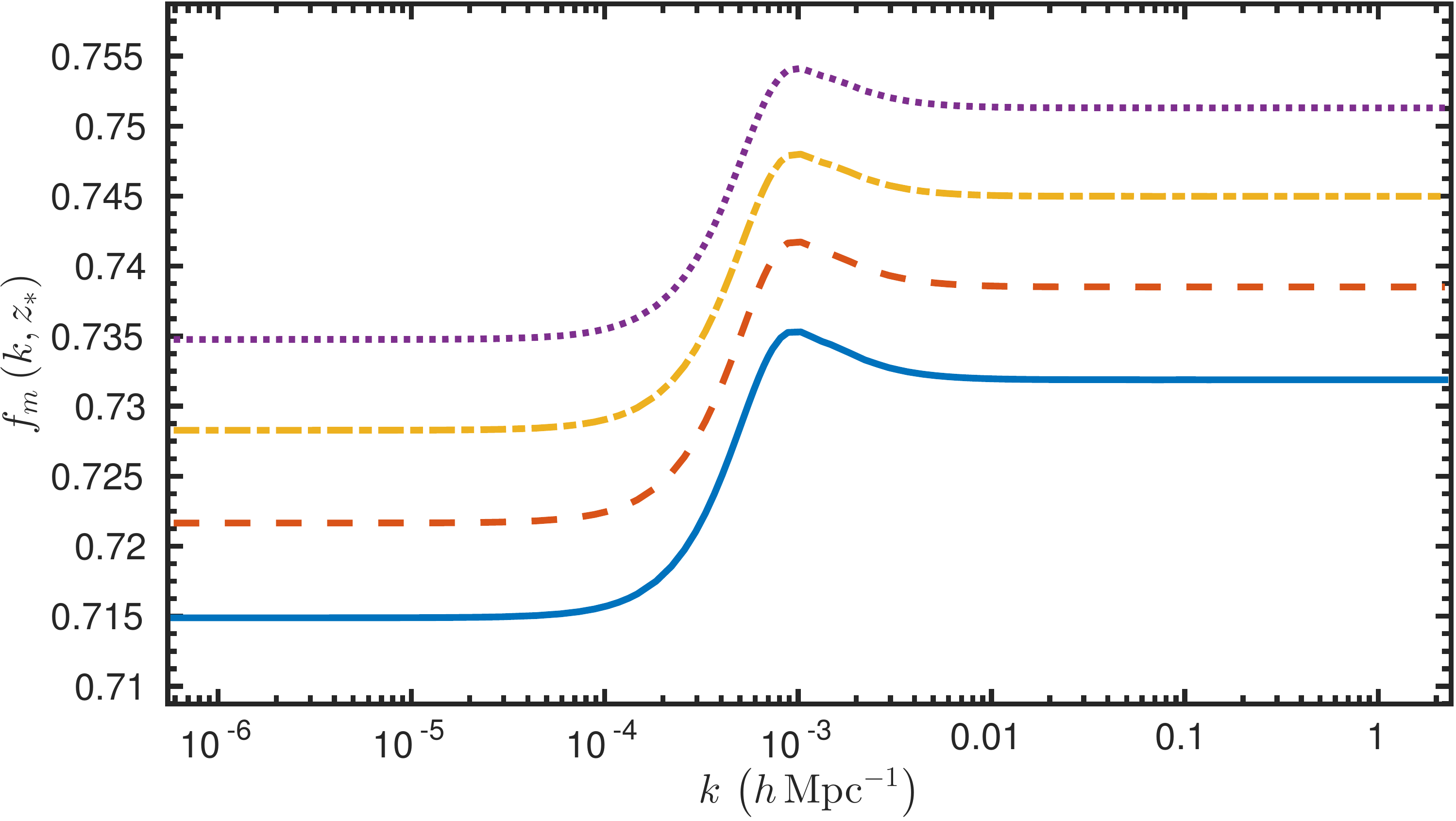}
  \includegraphics[width=0.48\textwidth]{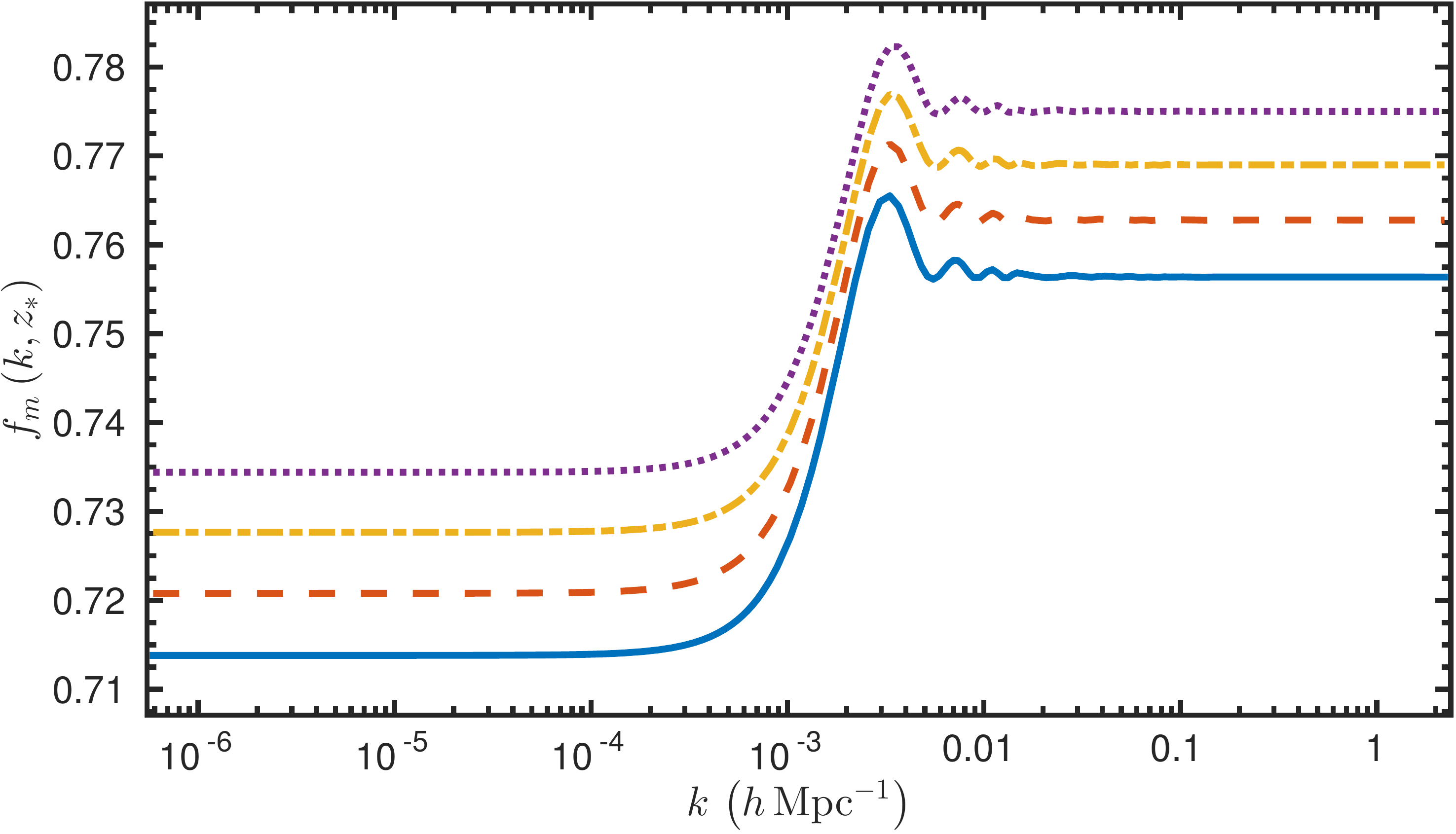}
  \includegraphics[width=0.48\textwidth]{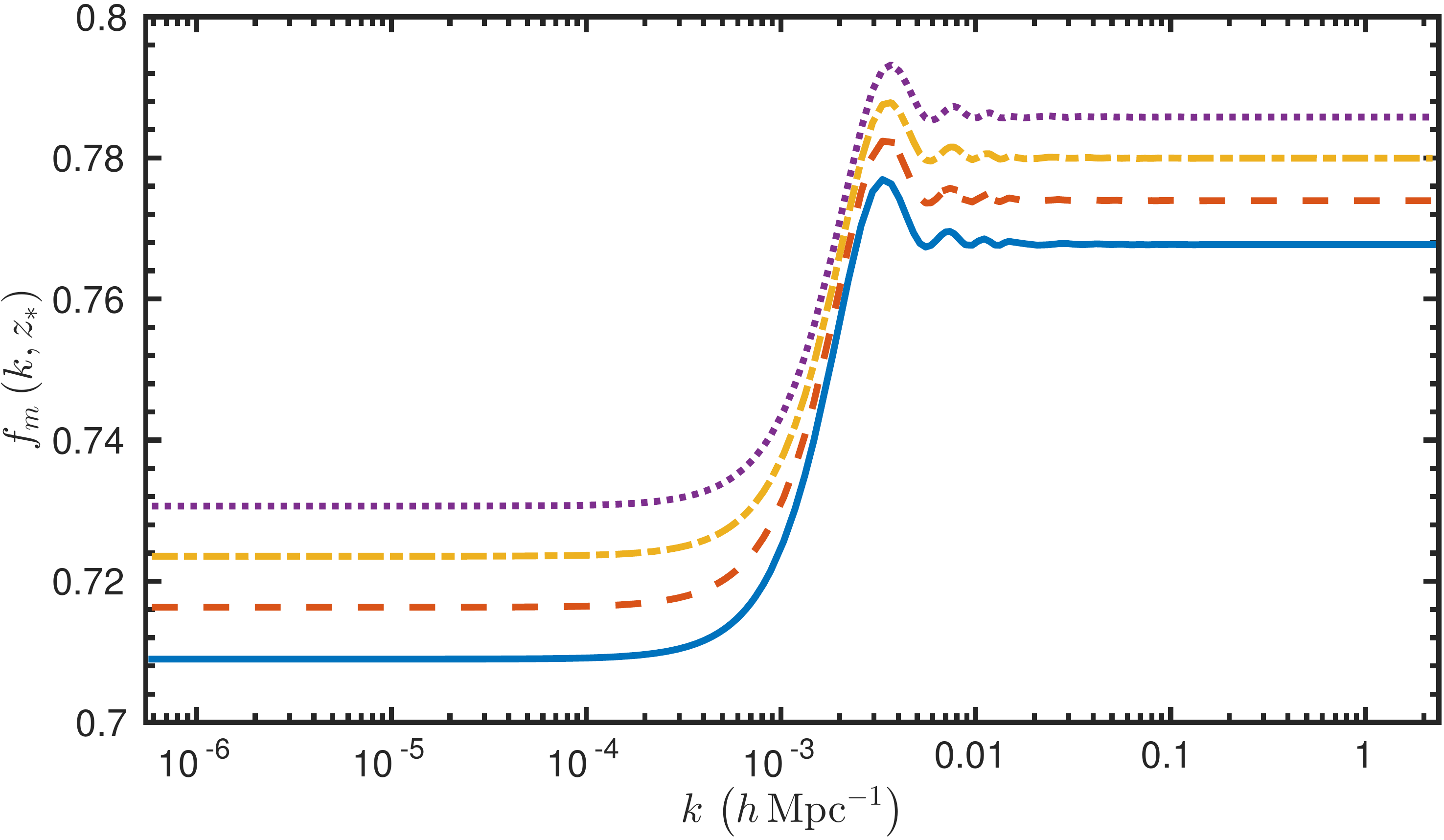}
\caption{These figures show the matter growth rate function $f_m\left(k,\,z_\ast\right)$ as a function of the wave number $k$ in $h\,\text{Mpc}^{-1}$ at the redshifts $z_\ast=0.50,\,0.52,\,0.54,\,0.56$. The uncoupled case is shown in the top left plot, the top right plot is the conformal case, the lower left plot is the disformal case, and the lower right plot is the mixed case. The model parameters are the same as in Fig. \ref{fig:fs8_HHo}.  }  
\label{fig:f_m_k}
\end{figure}

An important feature in Fig. \ref{fig:fs8_HHo} is the turnaround location, which is easily distinguishable for each different model depicted in this figure. This turning point in each curve comes from the fact that as the models enter the accelerating epoch, the growth rate is suppressed with respect to its value in the matter dominated era. Although the expansion history of these models might not be a suitable discriminator, the growth history at late--times turns out to be more informative. For a given value of $H/H_0$, one can determine if the growth rate is enhanced or suppressed with respect to a specific model. Indeed, one can observe that conformally coupled models tend to give an enhanced growth rate with respect to the uncoupled case at all redshifts. On the other hand, models with a disformal coupling tend to suppress the growth of structure when the coupling is still not active, and when the late--time coupling starts to modify the cosmic evolution, the growth rate is enhanced, and overtakes the growth rates of the uncoupled and the conformal models (see also Ref. \cite{Jack}). One should also remark that a mixed model tends to be characterised by the largest growth rate as both couplings are contributing for this enhanced growth. This is an interesting feature of the disformal coupling, which distinguishes it from the rest.

\begin{figure}[t!]
\centering
  \includegraphics[width=0.48\textwidth]{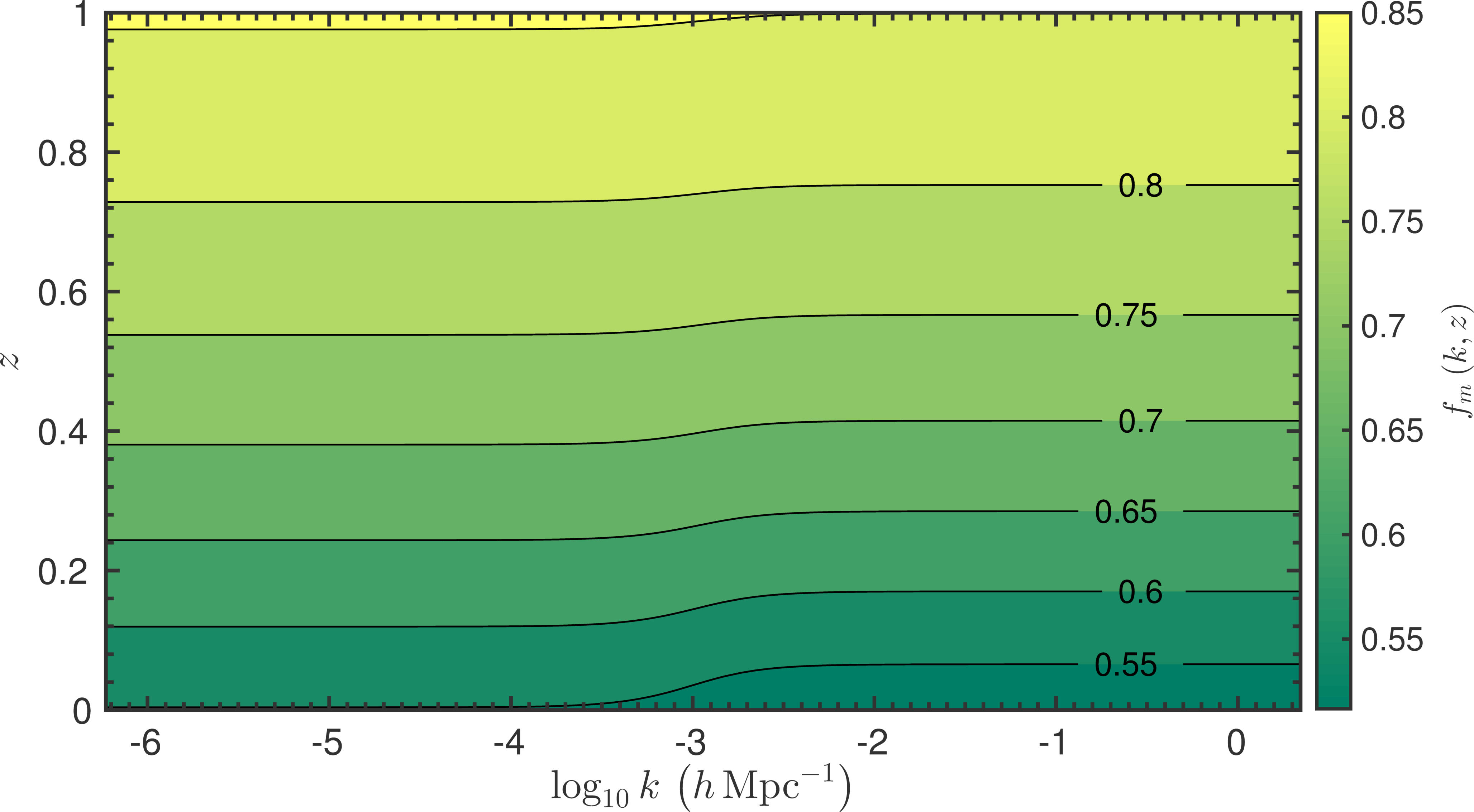}
  \includegraphics[width=0.48\textwidth]{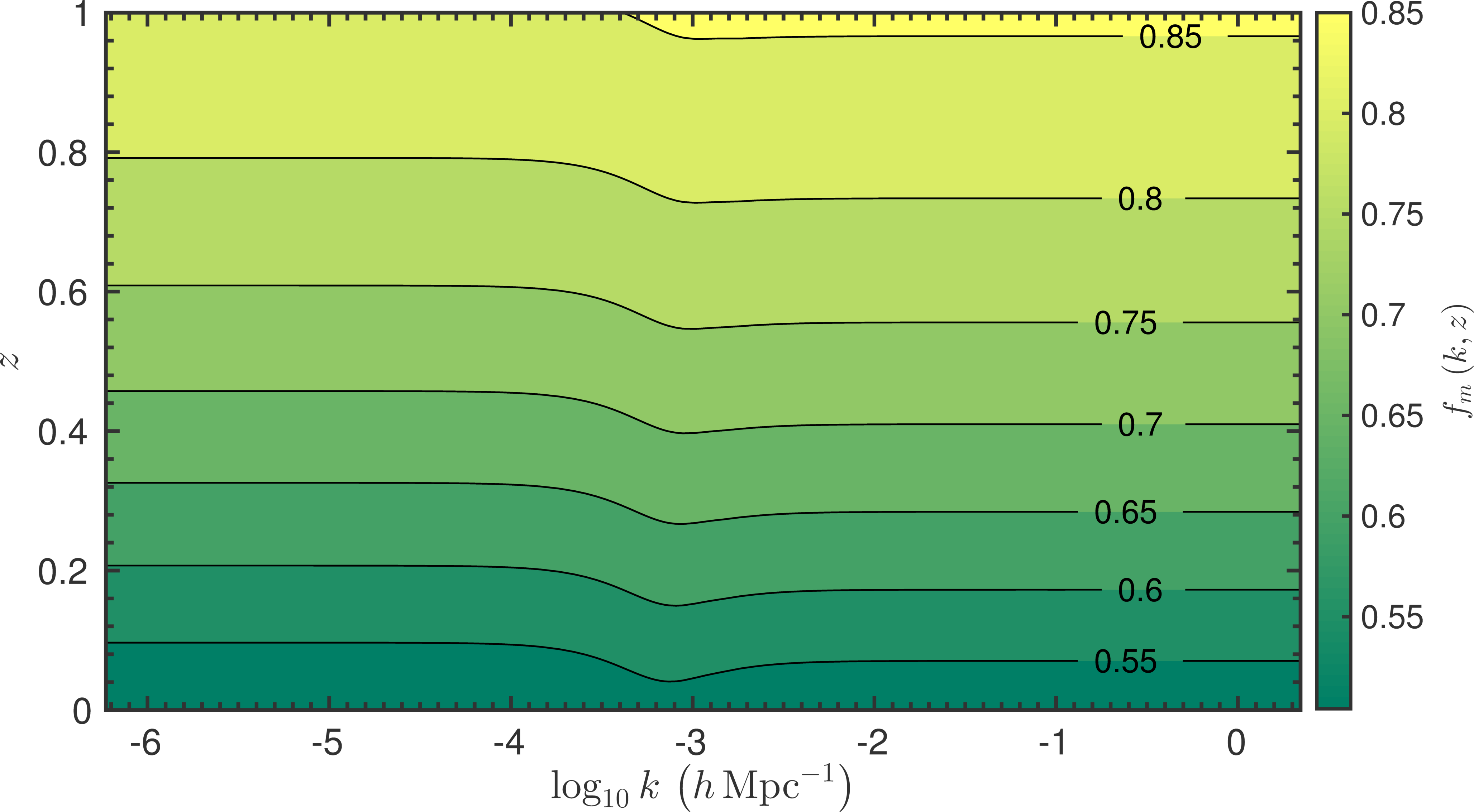}
  \includegraphics[width=0.48\textwidth]{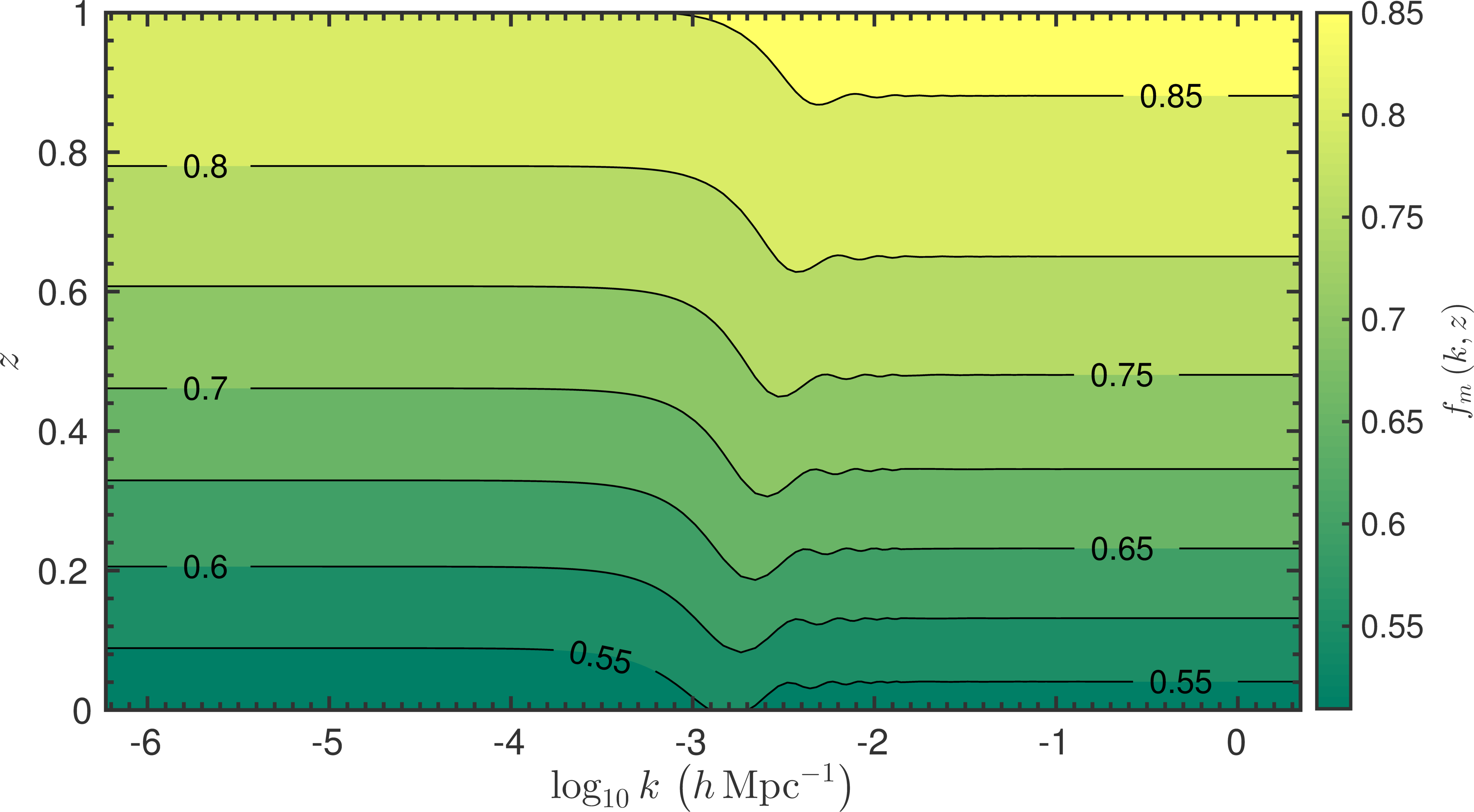}
  \includegraphics[width=0.48\textwidth]{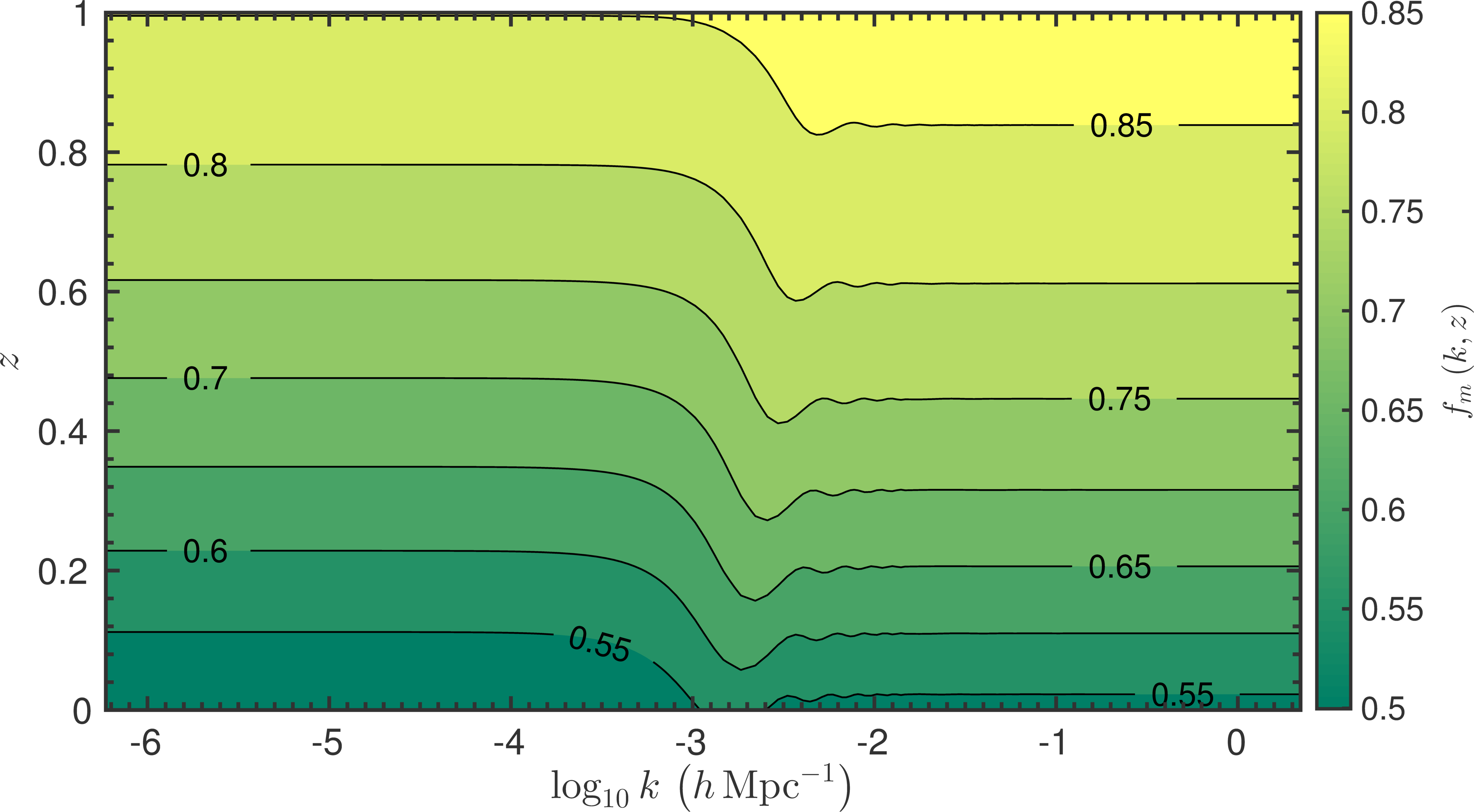}
\caption{These figures show the contour lines of the matter growth rate function $f_m\left(k,\,z\right)$ as a function of the wave number $\log_{10}k$ in $h\,\text{Mpc}^{-1}$ and redshift $z$. The uncoupled case is shown in the top left plot, the top right plot is the conformal case, the lower left plot is the disformal case, and the lower right plot is the mixed case. The model parameters are the same as in Fig. \ref{fig:fs8_HHo}.  }  
\label{fig:f_m_k_z_low}
\end{figure}

We now consider the evolution of $f_m(k,z)$ as a function of the wave number $k$, which covers both the large--scales as well as the small--scales, at some particular redshifts. We present the plots of this wave number evolution in Fig. \ref{fig:f_m_k}, in which we illustrate four different models, including standard quintessence together with the coupled models. As expected, the growth rate in the standard quintessence model, can be regarded as being (nearly) $k$--independent for the whole range of values being considered in this plot. On the other hand, coupled models are characterised by an enhancement in the growth rate function on the small--scales when compared with the large--scales. In the conformal model, this is a well--known characteristic (see for example Refs. \cite{Xia:2009zzb,Amendola:2011ie}) which is easily observed from the increase in power in the matter power spectrum on small--scales. In this scenario, the increase in growth rate on small--scales is a result of the fact that due to the coupling there is an increase in the DM fraction in the past when compared with that in the uncoupled scenario, leading to an earlier matter radiation equality. This implies that the wavelengths of the perturbations that enter during the radiation dominated era are shorter, and therefore the turnaround of the matter power spectrum moves to smaller scales and the small--scale amplitude of the matter power spectrum is boosted. Another feature in the matter power spectrum is the change in location and amplitude of the baryon acoustic oscillations peaks imprinted on the matter power spectrum itself.

In interacting DE models which include a disformal coupling, the increase in the matter growth rate function on small--scales is mainly due to the additional attractive force between the DM particles as a result of their coupling. In section \ref{sec:Newtonian_limit}, we will find that on these scales, the attractive force between the coupled DM particles is enhanced, leading to an increase in the growth rate function. This also holds for the conformally coupled models, although disformal couplings tend to be associated with a larger enhancement of this additional force. As a consequence of this enhancement in the growth of structure on small--scales, the $\sigma_8(z=0)$ value is also expected to increase in these models. It would be worth exploring the model parameters by, for example, a Markov chain Monte Carlo exploration in light of the growth of structure constraints, although such work is postponed for future work \cite{Mifsudjm:2017}. 

\begin{figure}[t!]
\centering
  \includegraphics[width=0.48\textwidth]{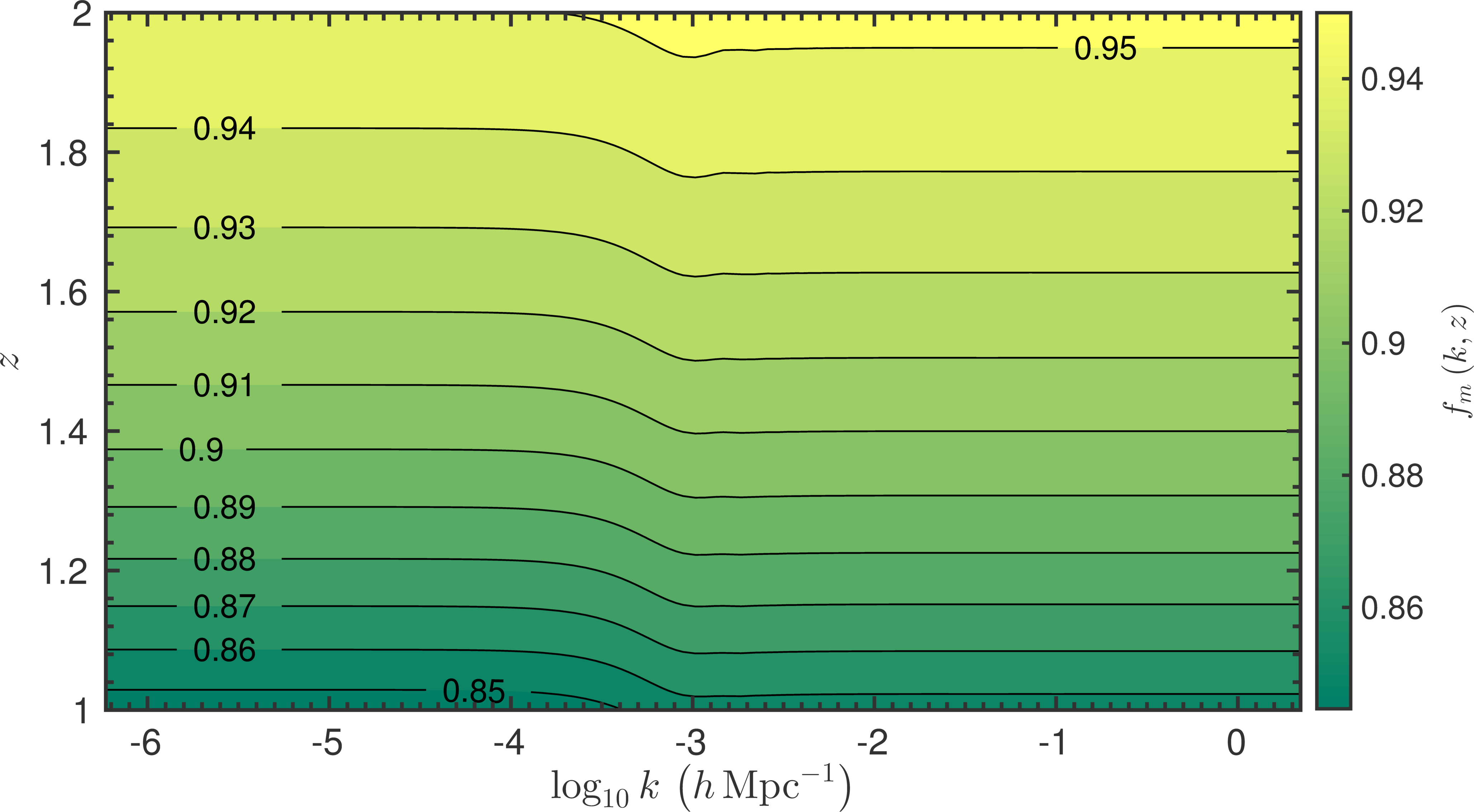}
  \includegraphics[width=0.48\textwidth]{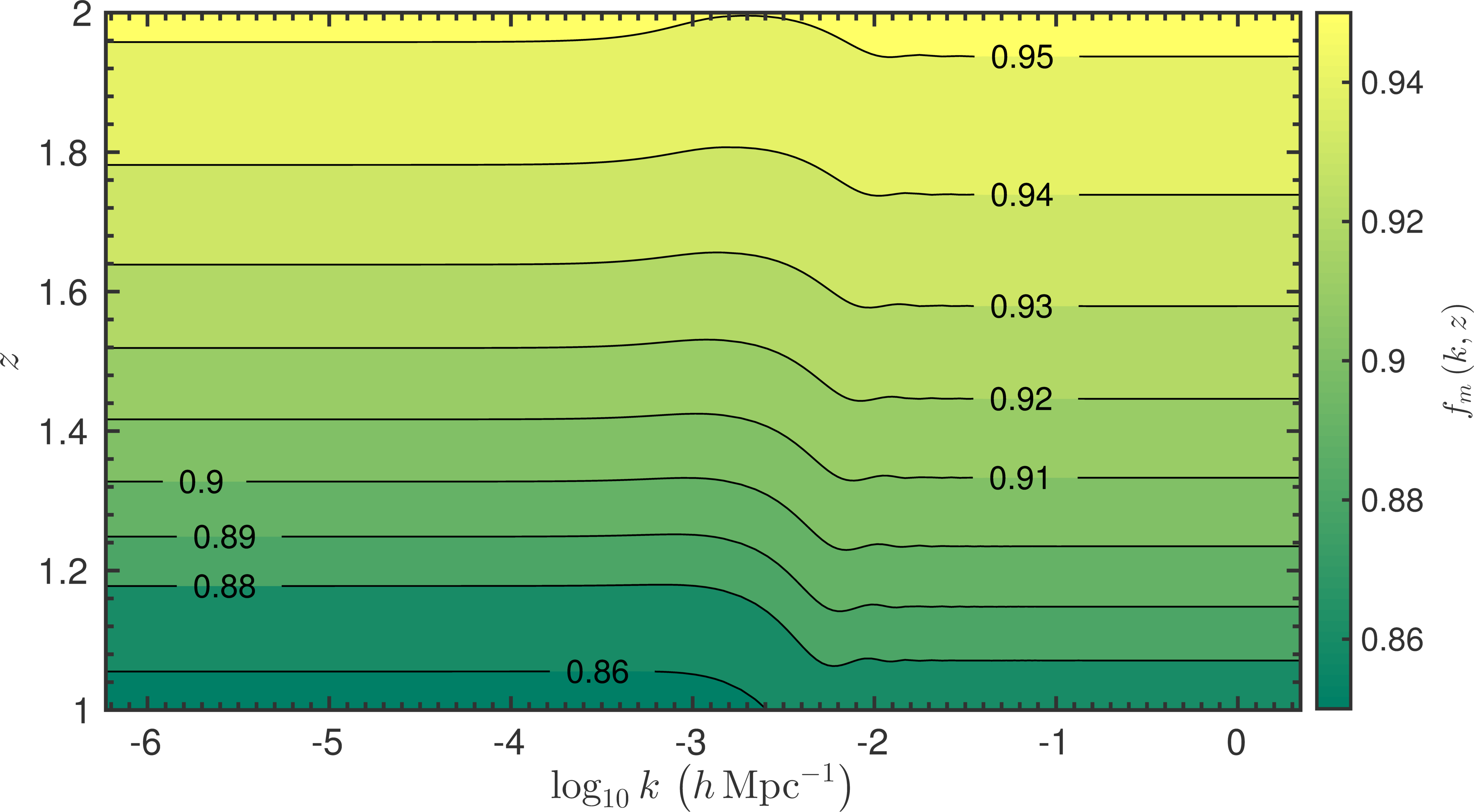}
  \includegraphics[width=0.48\textwidth]{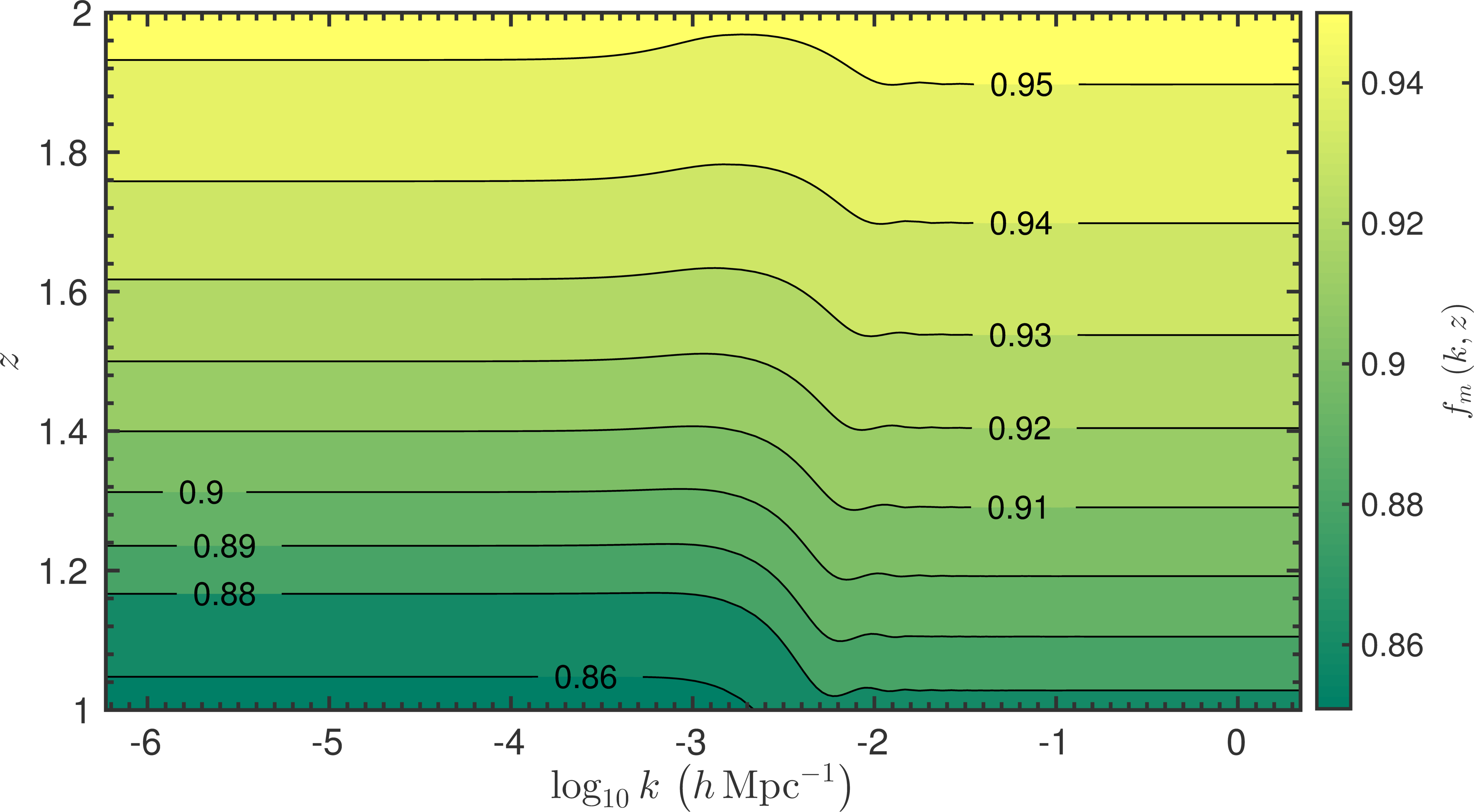}
\caption{These figures show the contour lines of the matter growth rate function $f_m\left(k,\,z\right)$ as a function of the wave number $\log_{10}k$ in $h\,\text{Mpc}^{-1}$ and redshift $z$. The top left plot is the conformal case, the top right plot is the disformal case, and the lower plot is the mixed case. The model parameters are the same as in Fig. \ref{fig:fs8_HHo}.  }  
\label{fig:f_m_k_z_high}
\end{figure}

In the presence of a disformal coupling, the matter growth rate function will be characterised by distinctive intermediate--scales and time--dependent damped oscillations attributed to the dynamics of the coupling function itself. These peculiar features only occur when one considers the disformal coupling, since these are not observed in standard quintessence or in conformally coupled models. The oscillations in the matter growth rate function are present when the disformal coupling starts to play an important role in the cosmic evolution, and thus we expect these oscillations to be negligible at higher redshifts. Indeed, this is what happens, as clearly shown in Fig. \ref{fig:f_m_k_z_low} and Fig. \ref{fig:f_m_k_z_high}, in which the oscillatory features are clearly visible at $z<1$, losing their significance even at $z\simeq1.5$. Also, from these contour plots, one can see that a disformal coupling induces a slight scale--dependence on the growth rate function. This is expected due to the $k$--dependence of the perturbed coupling function $\delta Q$ given in Eq. (\ref{delta_Q}). Such time--dependent and scale--dependent characteristics in the matter growth rate function could be probed by upcoming cosmological surveys, including emission--line--galaxy surveys together with intensity mapping experiments \cite{vanHaarlem:2013dsa}, measuring the scale--dependence of the matter power spectrum at several cosmic times \cite{Hernandez:2016xci}.

%
\section{The small--scale limit of perturbations}
\label{sec:Newtonian_limit}
We now discuss the Newtonian limit of the generic perturbation equations presented in Appendix \ref{sec:Perturbations} for a coupled barotropic pressureless fluid scenario. For this analysis, we neglect the anisotropic stress contribution in the field equations, leading to $\Psi=\Phi$. In the small--scale limit $\hat{\lambda}=\mathcal{H}/k\ll1$, the evolution equations of the gravitational potential $\Phi$, and its conformal time derivative $\Phi^\prime$, reduce to the following
\begin{align}
\Phi&\simeq-\frac{\hat{\lambda}^2}{2}\left[\frac{\kappa^2}{\mathcal{H}^2}\left(3\mathcal{H}\phi^\prime\delta\phi+\phi^\prime\delta\phi^\prime+a^2V_{,\phi}\delta\phi\right)+3\sum_{i=b,r,c} \Omega_i\delta_i\right]\;,\label{Poisson1}\\
\Phi^\prime&\simeq\frac{1}{2}\left(\kappa^2\phi^\prime\delta\phi-2\mathcal{H}\Phi\right)\;.
\end{align}
\begin{figure}[t!]
\centering
  \includegraphics[width=0.49\textwidth]{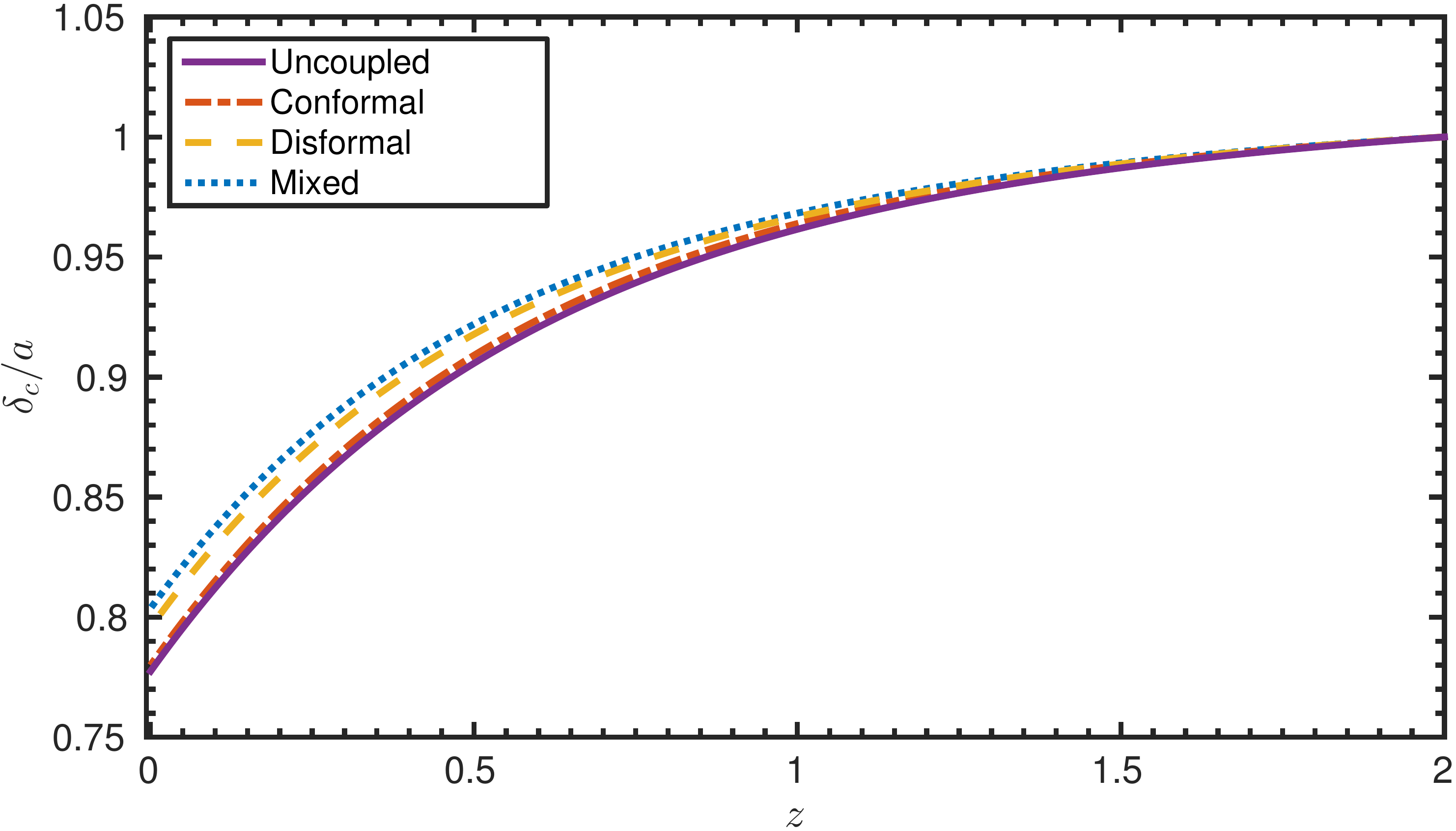}
  \includegraphics[width=0.49\textwidth]{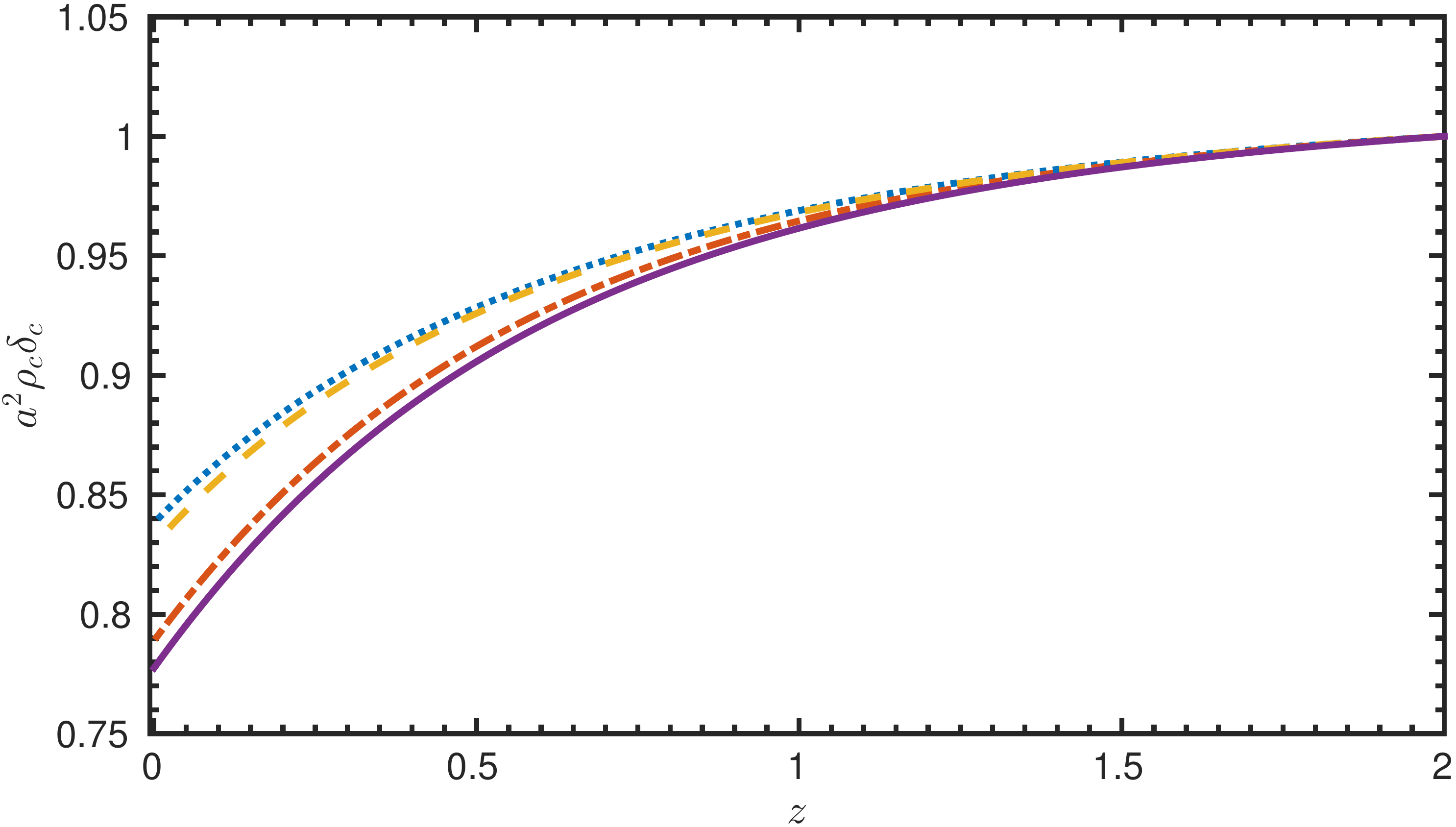} 
 \caption{These figures show the redshift evolution of the normalized DM growth rate $\delta_c/a$ (left), and the normalized combination of $a^2\rho_c\delta_c$ (right) appearing in the Poisson equation, at wave number $k=0.1\,h\,\text{Mpc}^{-1}$. All model parameters are the same as those used in Fig. \ref{fig:fs8_HHo}.}  
\label{fig:sub_horizon}
\end{figure}
Moreover, in this small--scale limit, the evolution of the perturbed scalar field is now governed by the equation
\begin{equation}\label{KG-small-scale}
\delta\phi^{\prime\prime}+2\mathcal{H}\delta\phi^\prime+\hat{\lambda}^{-2}\mathcal{H}^2\delta\phi\simeq a^2\delta Q\;,
\end{equation}
in which we have neglected terms proportional to $\Phi\,\big(\sim\hat{\lambda}^2\big)$. Furthermore, we assumed that the term proportional to ${\phi^\prime}^2$ is much less than $\hat{\lambda}^{-2}$, and that the potential is flat enough so that the $V_{,\phi\phi}$ term is negligible with respect to $\hat{\lambda}^{-2}$. The homogeneous solution of Eq. (\ref{KG-small-scale}) averages out to a zero contribution to the perturbed scalar field solution in the very small--scale limit, leaving only the inhomogeneous solution, which, on averaging over the oscillations and neglecting the contributions from $\delta\phi^{\prime\prime}$ and $\delta\phi^{\prime}$ (this can be further checked a posteriori), is found to be approximately equal to:
\begin{equation}
\delta\phi\simeq\hat{\lambda}^2a^2\mathcal{H}^{-2}\delta Q\;.
\end{equation}
In this limit, the perturbation of the coupling function, defined by Eq. (\ref{delta_Q}), simplifies significantly to \cite{Zumalacarregui:2012us, Jack} 
\begin{equation}
\delta Q\simeq Q\delta_c\;.
\end{equation}
Since $\delta\phi$ is of the order of $\hat{\lambda}^2$, Eq. (\ref{Poisson1}) reduces to the standard Poisson equation 
\begin{equation}\label{Newt_limit_Poisson}
\Phi\simeq-\frac{3}{2}\hat{\lambda}^2\sum_{i=b,r,c}\Omega_i\delta_i\;.
\end{equation}
From the term $a^2\rho_c\delta_c$ appearing on the right hand side of Eq. (\ref{Newt_limit_Poisson}), one can determine if this quantity changes drastically in these interacting DE models, which would then lead to a time-evolving gravitational potential. From Fig. \ref{fig:sub_horizon}, we can see that although the growth rates at late--times can differ from one another (especially when large couplings are considered), the combination $a^2\rho_c\delta_c$ does not change appreciably, thus leaving a small imprint of the ISW effect on the CMB temperature power spectrum, in agreement with the results obtained in section \ref{sec:ISW_DE_models}. This is an unusual behaviour of these models, since normally cosmological models with different growth history give rise to a distinguishable ISW effect \cite{Clemson:2011an}. 

Furthermore, the evolution of the coupled pressureless fluid density contrast is now governed by the differential equation
\begin{equation}\label{density_contrast}
\delta_c^{\prime\prime} + \mathcal{H}_{\text{eff}}\,\delta_c^\prime-\frac{3}{2}\mathcal{H}^2\,\frac{G_{\text{eff}}}{G}\,\Omega_c\delta_c = \frac{3}{2}\mathcal{H}^2\left(\Omega_b\delta_b + \Omega_r\delta_r\right)\;.
\end{equation}
Hence, the coupled fluid perturbations experience effectively different values of $\mathcal{H}$ and $G$ due to the interaction \cite{Jack},
\begin{equation}
\frac{\mathcal{H}_{\text{eff}}}{\mathcal{H}}=1-\frac{1}{\mathcal{H}}\frac{Q}{\rho_c}\phi^\prime\;,\;\;\;\; \frac{G_{\text{eff}}}{G}=1+\frac{2}{\kappa^2}\frac{Q^2}{\rho_c^2}\;.
\end{equation}
Thus, the introduction of a coupling between the DE scalar field and DM, induces a modification in the damping term together with an amplification of Newton's gravitational constant in Eq. (\ref{density_contrast}). Moreover, the added contribution in the effective gravitational constant is independent from the sign of the coupling function. 

Since baryons satisfy the standard uncoupled equation for the evolution of the baryon density contrast, we expect that there will be a bias between baryons and coupled DM. We study this in the DM dominated scenario, $|\Omega_c\delta_c|\gg|\Omega_b\delta_b|\gg|\Omega_r\delta_r|$, and define a constant bias $b$, by $\delta_b=b\,\delta_c$. We can easily determine the bias by writing Eq. (\ref{density_contrast}) and a similar one for baryons (in which we also neglect the term proportional to its sound speed) in terms of the coupled DM growth parameter $f_c=d\ln\delta_c/dN$, where $N=\ln a$. Indeed, we find that the growth rate equations of baryons and coupled DM reduce to
\begin{align}
\frac{df_c}{dN}+f_c^2+\frac{1}{2}\left(1-3w_{\text{eff}}\right)f_c-\frac{3}{2}\frac{\Omega_c}{b}&=0\label{baryon_growth}\;,\\
\frac{df_c}{dN}+f_c^2+\frac{1}{2}\left(1-3w_{\text{eff}}-2\frac{Q}{\rho_c}\frac{d\phi}{dN}\right)f_c-\frac{3}{2}\frac{G_{\text{eff}}}{G}\Omega_c&=0\label{CDM_growth}\;,
\end{align}
respectively, where we defined a total effective equation of state, as customary called in dynamical systems analysis (not to be confused with the previously defined effective equations of state), which characterises the expansion rate as
\begin{equation}\label{effective_EOS}
\frac{1}{\mathcal{H}}\frac{d\mathcal{H}}{dN}=-\frac{1}{2}\left(1+3w_{\text{eff}}\right)\;.
\end{equation}
From Eq. (\ref{baryon_growth}) and Eq. (\ref{CDM_growth}), one arrives to a simplified expression for the bias
\begin{equation}
b=\frac{3\Omega_c}{2\frac{Q}{\rho_c}\frac{d\phi}{dN}f_c+3\frac{G_{\text{eff}}}{G}\Omega_c}\;.
\end{equation}
Indeed, as a result of unequal couplings of these pressureless species, a time-dependent bias develops between them.
\subsection{Analytical solutions in interacting dark energy models}
\label{sec:analytical_sol}
We will now briefly discuss some analytical solutions of Eq. (\ref{density_contrast}) at four particular coupled fixed points in the DM dominated era. The effective equation of state defined in Eq. (\ref{effective_EOS}) is constant at these fixed points, thus the scale factor evolves as $a\sim\tau^{2/(1+3w_{\text{eff}})}$. The fixed points of the coupled models that we are considering in this work have been extensively discussed in Ref. \cite{vandeBruck:2016jgg}, which we now follow. For this section only, we shall consider the following coupling and scalar field potential functions
\begin{equation}
C(\phi)=e^{2\alpha\kappa\phi}\;,\;\;\;\;D(\phi)=D_M^4 e^{2\left(\alpha+\beta\right)\kappa\phi}\;,\;\;\;\;V(\phi)=V_0^4 e^{-\lambda\kappa\phi}\;,
\end{equation}
where we recall that $\alpha,\,D_M,\,\beta,\,V_0,$ and $\lambda$ are constants. At any fixed point, one can conveniently write Eq. (\ref{density_contrast}) as follows
\begin{equation}
\frac{d^2\delta_c}{dN^2} + \xi_1\frac{d\delta_c}{dN} + \xi_2\delta_c = 0\;,
\end{equation}
where $\xi_1$ and $\xi_2$ are both constants which depend on the phase-space coordinates of that particular fixed point. Thus, the solution of the coupled DM density contrast is
\begin{equation}
\delta_c = c_{+} a^{m_{+}} + c_{-} a^{m_{-}}\;,\;\;\;\;m_{\pm}=\frac{1}{2}\left(-\xi_1\pm\sqrt{\xi_1^2-4\xi_2}~\right)\;,
\end{equation}
where $c_{\pm}$ are integration constants. Moreover, from Eq. (\ref{Newt_limit_Poisson}) we find that at these fixed points $\Phi\sim a^{-1-3w_{\text{eff}}+m_{\pm}}$.
\subsubsection*{Disformal fixed points}
We shall consider the two disformal fixed points $(3)_{(d)}$ and $(4)_{(d)}$, reported in Ref. \cite{vandeBruck:2016jgg}. For the fixed point $(3)_{(d)}$, which exists when $\beta\geq\sqrt{3/2}$, we find that this leads to a non--standard growth index
\begin{equation}
\begin{split}
m_{\pm} = \frac{1}{2}\Bigg\{ &-5 + 3\beta\left(2\beta-\sqrt{4\beta^2-6}\right) \\
&\pm\sqrt{1-2\beta\left\{\sqrt{4\beta^2-6} + \beta\left[1-2\beta\left(2\beta-\sqrt{4\beta^2-6}\right)\right]\right\} } ~\Bigg\}\;.
\end{split}
\end{equation}
Similarly, the disformal fixed point $(4)_{(d)}$, which exists for $\beta\leq-\sqrt{3/2}$, is characterised by the growth index
\begin{equation}
\begin{split}
m_{\pm} = \frac{1}{2}\Bigg\{ &-5 + 3\beta\left(2\beta+\sqrt{4\beta^2-6}\right)\\
& \pm\sqrt{1+2\beta\left\{\sqrt{4\beta^2-6} + \beta\left[-1+2\beta\left(2\beta+\sqrt{4\beta^2-6}\right)\right]\right\} } ~\Bigg\}\;.
\end{split}
\end{equation}
\begin{figure}[t!]
\centering
  \includegraphics[width=0.8\textwidth]{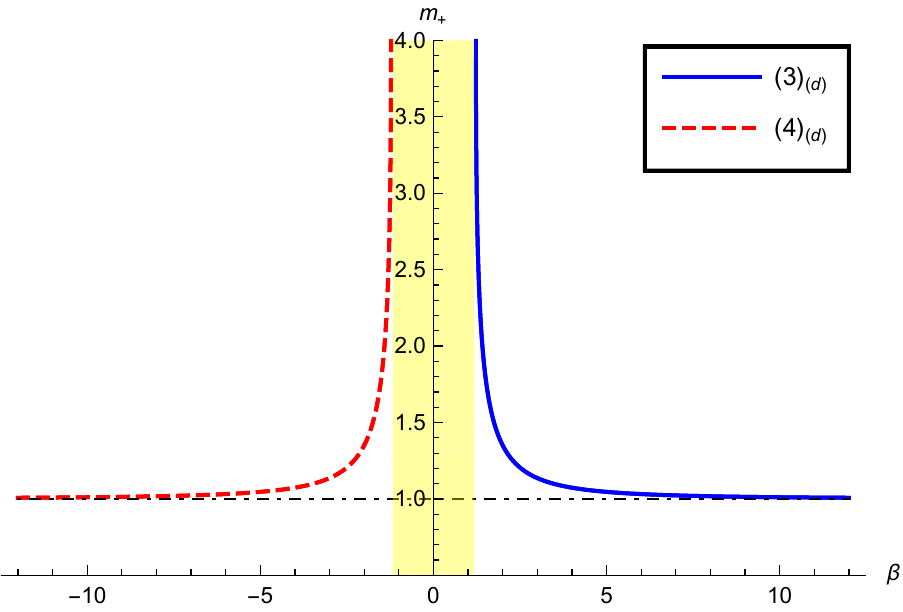}
 \caption{This is a plot showing the growth index $m_{+}$, as a function of the coupling parameter $\beta$, for the disformal fixed points $(3)_{(d)}$ and $(4)_{(d)}$. The shaded yellow region depicts the range of values of $-\sqrt{3/2}<\beta<\sqrt{3/2}$, at which both fixed points are not defined.} 
\label{fig:disformal}
\end{figure}
In Fig. \ref{fig:disformal}, we illustrate the growth index as a function of the coupling parameter $\beta$ for both disformal fixed points. At these fixed points, a non--standard growth index is only obtained for a restricted range of the parameter $\beta$. Moreover, we find that for the values of $\beta$ that we are considering, $\Phi$ is a constant to a very good approximation.
\subsubsection*{Conformal fixed points}
We will now cover the conformal scaling fixed point ($(8)_{(d)}$ in Ref. \cite{vandeBruck:2016jgg}), and another transient fixed point which appears in the DM dominated era giving rise to a scalar field matter dominated regime ($\phi$MDE) ($(6)_{(d)}$ in Ref. \cite{vandeBruck:2016jgg}). The latter fixed point is characterised by
\begin{equation}
m_{+} = 1+2\alpha^2\;,\;\;\;\;m_{-}=-\frac{3}{2}+\alpha^2\;,
\end{equation} 
leading to an enhanced growth rate of coupled DM when compared with the uncoupled scenario. Moreover, $\Phi$ is a constant at the $\phi$MDE growing mode solution. 

\begin{figure}[t!]
\centering
  \includegraphics[width=0.55\textwidth]{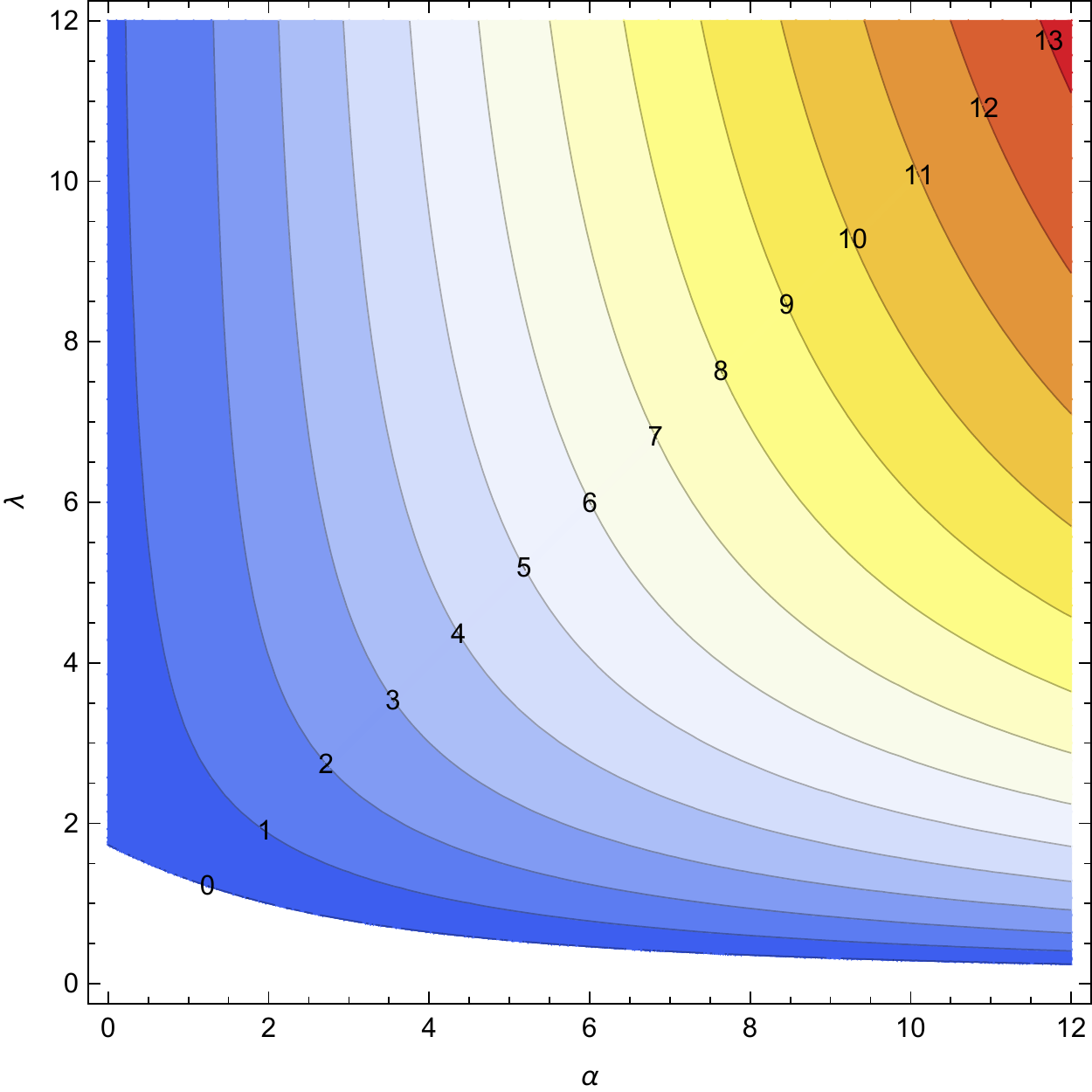}
 \caption{This is a contour plot of the growth index $m_{+}$, for the conformal scaling fixed point as a function of the conformal coupling parameter $\alpha$, and the slope of the exponential potential $\lambda$.} 
\label{fig:phiMDE}
\end{figure}

On the other hand, for the conformal scaling fixed point we have
\begin{equation}
m_{\pm}=\frac{1}{4}\left[-1+9w_{\text{eff}}\pm\sqrt{\left(1-9w_{\text{eff}}\right)^2 + 24\left(1-\Omega_\phi\right)\left(1+\frac{6w_{\text{eff}}^2}{\Omega_\phi+w_{\text{eff}}}\right)}~\right]\;,
\end{equation} 
where we used $\Omega_\phi=1-\Omega_c$. This growing mode solution gives rise to an enhanced growth of the DM perturbations, and to an anomalous ISW effect in the CMB power spectrum \cite{TocchiniValentini:2001ty}. In Fig. \ref{fig:phiMDE} we illustrate the growth index $m_{+}$, as a function of $\alpha$ and $\lambda$, where we have also used the relations
\begin{equation}
\Omega_\phi=\frac{3+\alpha(\alpha+\lambda)}{(\alpha+\lambda)^2}\;,\;\;\;\;w_{\text{eff}}=-\frac{\alpha}{\alpha+\lambda}\;.
\end{equation} 
%
%

\section{Conclusions}
\label{sec:conclusions}
Although the concordance model of cosmology is found to be in an excellent agreement with current cosmological observations, the theoretical framework is not fully satisfactory. Consequently, alternative scenarios have been extensively studied in the literature mostly to address the puzzling late--time accelerated expansion of the Universe.

In this work, we have considered a coupled quintessence model, in which the accelerated cosmic expansion is powered by the quintessence scalar field which is explicitly coupled to DM. In particular, the dark sector constituents were coupled via a conformal and a disformal coupling, whereas the baryonic and radiation sectors followed their standard cosmic evolution. The main aim of this paper was to study the cosmological implications of such a coupling, and to shed light on the characteristic signatures of the uncoupled, conformal, disformal, and mixed coupled models, thereby extending and clarifying the results of Ref. \cite{Jack}. 

The interaction between DE and DM can be viewed as an energy exchange mechanism between the two dark sector elements. This has been discussed while presenting the background evolution of the coupled models, in which one can easily observe that a disformal coupling is characterised by late--time modifications of the cosmic history, unlike the pure conformally coupled model. 

In order to study the implications of the dark sector coupling on the CMB temperature power spectrum, we have considered the multipole separation of the location of the peaks in the CMB temperature power spectrum together with the ISW effect. For the former, we presented an analytical approach which enabled us to look at all the different coupled models being considered in this paper. The deviations of the conformally coupled model from the $\Lambda$CDM model were found to be much larger than those in the mixed and the disformally coupled models. Indeed, we found that one is not able to distinguish between the concordance model and the coupled models which include a disformal coupling by only looking at the deviation of the CMB peak separation from the $\Lambda$CDM model. Moreover, we found that the discrepancy that arises from the ISW effect between the coupled quintessence models and the $\Lambda$CDM model, is not able to decipher the models from one another due to the uncertainty attributed with cosmic variance. 

We then considered the cosmological imprints on the growth of structure in interacting DE models. By plotting the expansion history against the growth history, we were able to clearly distinguish between the interacting DE models themselves along with the uncoupled model. We found that the coupling between DE and DM leads to an enhanced growth with respect to the uncoupled quintessence model, particularly in coupled models which include a disformal coupling. Furthermore, we discussed the matter growth rate function as a function of the wave number, which extends from the small--scales to the large--scales, as well as a function of the redshift. We found that the matter growth rate function is enhanced on small--scales as compared to large--scales in all interacting DE models. This observed enhancement was then studied via the small--scale approximation of the perturbation equations, in which we also discussed analytical solutions to the coupled DM density contrast at four specific fixed points.

Interestingly enough, disformal couplings were characterised by distinctive intermediate--scales and time--dependent damped oscillations in the matter growth rate function. This enabled us to further distinguish the coupled models making use of the disformal coupling from the rest. Forthcoming cosmological surveys should be able to measure the time--dependence and scale--dependence of the matter growth rate function, which could then provide additional constraints for these interacting DE models.

We conclude by briefly mentioning another possibility: the DM sector could consist of several species of DM, each with its own couplings to DE or species which are uncoupled. For example, it could be that there is a species of DM particles which is coupled to the DE scalar field in the way we discussed in this paper, and another species which is uncoupled. Depending on the abundance of the uncoupled species, the effects of the coupling discussed in this paper will be reduced. If the uncoupled DM species dominates, the growth rate (discussed in section \ref{sec:growth_history}) will be dominated by that species and the features observed in the matter growth rate function will be smaller. At the background level, the total DM fluid behaves like a single dark fluid coupled to DE, but with a reduced effective coupling. This situation is similar to the one discussed in Refs. \cite{Brookfield:2007au,Baldi:2014tja}. This would also address the concerns of Ref. \cite{DAmico:2016jbm} that a quantum field theory of a coupled DM species is hard to realise, unless DM consists (partly) of axions.

It remains to be seen whether and how the models studied here can be embedded in a more fundamental theory. First steps in this direction have been taken in Ref. \cite{Koivisto}. Clearly, more work needs to be done. 

\acknowledgments{We are grateful to Fernando Atrio-Barandela for indicating us Refs. \cite{Olivares:2008bx} and \cite{Wang:2016lxa}, as well as for useful correspondence. The work of CvdB is supported by the Lancaster-Manchester-Sheffield Consortium for Fundamental Physics under STFC Grant No. ST/L000520/1.}

\begin{subappendices}

\section*{Appendix A: Evolution of perturbations in interacting dark energy models}
\renewcommand{\theequation}{A.\arabic{equation}}
\renewcommand{\thesection}{A}
\refstepcounter{section}
\label{sec:Perturbations}
\noindent{We here present the equations governing the evolution of perturbations in our coupled models, in which we consider the perturbation equations for a generic interacting perfect fluid. The study of the growth of small perturbations about an FRW metric, given by the line element in Eq. (\ref{FRW}), is an asset in the understanding of the real Universe \cite{Bardeen:1980kt,Kodama:1985bj,Mukhanov:1990me,Ma:1995ey}. We will first discuss the relevant equations in the synchronous gauge, and we later derive the equations that govern the evolution of perturbations in the conformal Newtonian gauge. For the numerical solutions of our models, although the synchronous gauge equations were used throughout this paper, we have checked that the obtained results in both gauges agree with one another. The exact numerical solutions were computed in the Boltzmann code \texttt{CLASS}. }

\subsection*{Appendix A1: Synchronous gauge}
\renewcommand{\theequation}{A1.\arabic{equation}}
\renewcommand{\thesection}{A1}
\refstepcounter{section}
\label{sec:App_A1}
The line element in the synchronous gauge is given by
\begin{equation}
ds^2= a^2(\tau)\left[-d\tau^2 + \left(\delta_{ij}+h_{ij}\right) dx^i dx^j\right]\;,
\end{equation}
with metric perturbation $h_{ij}$. We adopt the convention of Ref. \cite{Ma:1995ey} and use the two metric perturbation fields $h$ and $\eta$ expressed in Fourier space $k$. In order to compute the first--order perturbed Einstein field equations $\delta\tensor{G}{^\mu _\nu}=8\pi G\sum\delta\tensor{T}{^\mu _\nu}$, we need the first--order perturbation of the zeroth--order energy--momentum tensor specified in Eq. (\ref{em_tensor}), leading to
\begin{equation}\label{em_perturb}
\delta\tensor{T}{^\mu_\nu}=\left(\delta\rho+\delta p\right)\bar{u}^\mu\bar{u}_\nu+\delta p\tensor{\delta}{^\mu_\nu}+(\rho+p)\left(\delta u^\mu\bar{u}_\nu+\bar{u}^\mu\delta u_\nu\right)+p\tensor{\Pi}{^\mu_\nu}\;,
\end{equation}  
where $\bar{u}^\mu$ is the zeroth--order four--velocity of the fluid, with $\delta u^\mu$ being its first--order perturbation. Moreover, $\Pi_{\mu\nu}$ is the traceless anisotropic stress tensor which characterises the difference between the perturbed fluid and a perfect fluid. The perturbations of the energy density $\delta\rho$, and pressure $\delta p$, are of the same order as the metric perturbations. The perturbed Einstein field equations reduce to the following set of coupled differential equations
\begin{align}
k^2\eta-\frac{1}{2}\mathcal{H}h^\prime &= -4\pi Ga^2\sum \delta\rho\;,\\
k^2\eta^\prime &= 4\pi Ga^2\sum\rho(1+w)\theta\;,\\
h^{\prime\prime}+2\mathcal{H}h^{\prime}-2k^2\eta &= -24\pi Ga^2\sum\delta p\;,\\
h^{\prime\prime}+6\eta^{\prime\prime}+2\mathcal{H}\left(h^{\prime}+6\eta^{\prime}\right)-2k^2\eta &= -24\pi Ga^2\sum\rho(1+w)\sigma\;,
\end{align}    
where the sum is over the DM, radiation, and DE fluids as explicitly written in Eq. (\ref{EFE}). The re--defined anisotropic stress perturbation $\sigma$, is related to the scalar part of the anisotropic stress tensor $\Pi$, as defined in Eq. (\ref{em_perturb}), by the relation $\sigma=2w\Pi/3(1+w)$. Moreover, the divergence of the fluid velocity is denoted by $\theta$. 

The perturbed continuity and Euler equations of the uncoupled baryonic and radiation (photons and massless neutrinos) sectors are governed by the standard first--order perturbation equations $\delta\tensor{T}{^\mu_\nu_;_\mu}=0$, which simplify to the following set of coupled differential equations
\begin{align}
\delta_i^\prime + 3\mathcal{H}\left(\frac{\delta p_i}{\delta\rho_i}-w_i\right)\delta_i &= -\left(1+w_i\right)\left(\theta_i+\frac{h^\prime}{2}\right)\;,\\
\theta_i^\prime+\left[\mathcal{H}\left(1-3w_i\right)+\frac{w_i^\prime}{1+w_i}\right]\theta_i &= \frac{\delta p_i}{\delta\rho_i}\frac{k^2\delta_i}{1+w_i}-k^2\sigma_i\;,
\end{align}
where $i=\{b,\;r\}$, and the density contrast is denoted by $\delta\equiv\delta\rho/\rho$. We recall that for the radiation sector $w_r=\delta p_r/\delta\rho_r=1/3$, and for baryons $w_b=\delta p_b/\delta\rho_b\ll 1$ with $\sigma_b=0$. The only non--negligible contribution of the shear stress comes from the radiation sector \cite{Ma:1995ey}, which we include in our numerical solutions.

For the coupled fluid, the conservation equation is modified according to Eq. (\ref{cons}), leading to the following perturbed continuity and Euler equations
\begin{align}
\delta_c^\prime + 3\mathcal{H}\left(\frac{\delta p_c}{\delta\rho_c}-w_c\right)\delta_c &= -\left(1+w_c\right)\left(\theta_c+\frac{h^\prime}{2}\right) +\frac{Q}{\rho_c}\phi^\prime\delta_c-\frac{Q}{\rho_c}\delta\phi^\prime-\frac{\phi^\prime}{\rho_c}\delta Q\;,\\
\theta_c^\prime+\left[\mathcal{H}\left(1-3w_c\right)+\frac{w_c^\prime}{1+w_c}\right]\theta_c &= \frac{\delta p_c}{\delta\rho_c}\frac{k^2\delta_c}{1+w_c}+\frac{Q}{\rho_c}\phi^\prime\theta_c-\frac{Q}{\rho_c(1+w_c)}k^2\delta\phi\;,
\end{align}
which are valid for a coupled shear--free fluid with equation of state $w_c$. The perturbation of the coupled DE scalar field is denoted by $\delta\phi$, and its evolution is governed by the following perturbed Klein--Gordon equation
\begin{equation}
\delta\phi^{\prime\prime}+2\mathcal{H}\delta\phi^\prime+\left(a^2 V_{,\phi\phi}+k^2\right)\delta\phi+\frac{h^\prime}{2}\phi^\prime=a^2\delta Q\;.
\end{equation}
The corresponding perturbation of the coupling function $Q$ is given by
\begin{equation}
\delta Q=-\frac{\rho_c}{a^2C+D\left(a^2\rho_c-{\phi^\prime}^2\right)}\left(\mathfrak{B}_1\delta_c+\mathfrak{B}_2h^\prime+\mathfrak{B}_3\delta\phi^\prime+\mathfrak{B}_4\delta\phi\right)\;,
\end{equation}
where
\begin{align}
\mathfrak{B}_1=&\frac{1}{2}a^2C_{,\phi}\left(1-3\frac{\delta p_c}{\delta\rho_c}\right)-3\mathcal{H}D\left(1+\frac{\delta p_c}{\delta\rho_c}\right)\phi^\prime-a^2D\left(V_{,\phi}-Q\right)-D{\phi^\prime}^2\left(\frac{C_{,\phi}}{C}-\frac{D_{,\phi}}{2D}\right)\;,\\
\mathfrak{B}_2=&-\frac{1}{2}D\phi^\prime\left(1+w_c\right)\;,\\
\mathfrak{B}_3=&-3\mathcal{H}D\left(1+w_c\right)-2D\phi^\prime\left(\frac{Q}{\rho_c}+\frac{C_{,\phi}}{C}-\frac{D_{,\phi}}{2D}\right)\;,\\
\mathfrak{B}_4=&\frac{1}{2}a^2C_{,\phi\phi}\left(1-3w_c\right)-\left(1+w_c\right)k^2D-a^2DV_{,\phi\phi}-a^2D_{,\phi}V_{,\phi}-3\mathcal{H}D_{,\phi}\left(1+w_c\right)\phi^\prime\nonumber\\
&-D{\phi^\prime}^2\left(\frac{C_{,\phi\phi}}{C}-\frac{C_{,\phi}^2}{C^2}+\frac{C_{,\phi}D_{,\phi}}{CD}-\frac{1}{2}\frac{D_{,\phi\phi}}{D}\right)+\frac{Q}{\rho_c}\left(a^2C_{,\phi}+a^2D_{,\phi}\rho_c-D_{,\phi}{\phi^\prime}^2\right)\;.
\end{align}
For the pure disformal scenario, i.e. when $C(\phi)=1$, the perturbation of $Q$ simplifies to the following equation
\begin{equation}
\begin{split}
\delta Q^{(d)}=&\left[\left(a^2-D{\phi^\prime}^2\right)\frac{Q}{\zeta}-\frac{3\mathcal{H}D\rho_c}{\zeta}\left(w_c-\frac{\delta p_c}{\delta\rho_c}\right)\phi^\prime\right]\delta_c+\frac{D\phi^\prime\rho_c}{\zeta}\left(1+w_c\right)\frac{h^\prime}{2}\\
&-\frac{\rho_c}{\zeta^2}\delta\phi^\prime\Big\{a^2D_{,\phi}\phi^\prime\left(1+D\rho_c\right)-D^2\left[2a^2V_{,\phi}\phi^\prime+3\mathcal{H}\left(1+w_c\right)\left(a^2\rho_c+{\phi^\prime}^2\right)\right]\Big.\\
&\Big.\;\;\;\;\;\;\;\;\;\;\;\;\;\;\;-3a^2\mathcal{H}D\left(1+w_c\right)\Big\}\\
&+\delta\phi\left\{k^2\frac{D\rho_c}{\zeta}\left(1+w_c\right)+\frac{\rho_c}{2\zeta}\left(2a^2DV_{,\phi\phi}-D_{,\phi\phi}{\phi^\prime}^2\right)\right.\\
&\left.\;\;\;\;\;\;\;\;\;\;+\frac{\rho_c}{2\zeta^2}\left[2a^2D_{,\phi}\left(a^2V_{,\phi}+3\mathcal{H}\phi^\prime\left(1+w_c\right)\right)+D_{,\phi}^2{\phi^\prime}^2\left(a^2\rho_c-{\phi^\prime}^2\right)\right]\right\}\;,
\end{split}
\end{equation}
where we define $\zeta=a^2+D\big(a^2\rho_c-{\phi^\prime}^2\big)$. In the absence of a disformal coupling the above perturbation equations simplify considerably. Indeed, in the pure conformal case (see also Ref. \cite{Amendola:2003wa}), the perturbed continuity and Euler equations for a generic coupled fluid reduce to
\begin{align}
\delta_c^\prime& + 3\left(\frac{\delta p_c}{\delta\rho_c}-w_c\right)\left(\mathcal{H}+\frac{1}{2}\left(\ln C\right)_{,\phi}\phi^\prime\right)\delta_c\nonumber\\
&=-\left(1+w_c\right)\left(\theta_c+\frac{h^\prime}{2}\right)+\frac{1}{2}\left(1-3w_c\right)\left[\left(\ln C\right)_{,\phi}\delta\phi^\prime+\left(\ln C\right)_{,\phi\phi}\phi^\prime\delta\phi\right]\;,
\\ \theta_c^\prime&+ \left[\mathcal{H}\left(1-3w_c\right)+\frac{w_c^\prime}{1+w_c}+\frac{1}{2}\left(\ln C\right)_{,\phi}\left(1-3w_c\right)\phi^\prime\right]\theta_c\nonumber\\
&=k^2\left[\frac{\delta p_c}{\delta\rho_c}\frac{\delta_c}{1+w_c}+\frac{1}{2}\left(\ln C\right)_{,\phi}\left(\frac{1-3w_c}{1+w_c}\right)\delta\phi\right]\;,
\end{align} 
and the perturbed Klein--Gordon equation reduces to
\begin{equation}\label{KG_conformal}
\begin{split}
\delta\phi^{\prime\prime}&+2\mathcal{H}\delta\phi^\prime+\left[k^2+a^2V_{,\phi\phi}+\frac{1}{2}a^2\rho_c\left(1-3w_c\right)\left(\ln C\right)_{,\phi\phi}\right]\delta\phi\\
&=-\frac{1}{2}h^\prime\phi^\prime-\frac{1}{2}a^2\rho_c\left(\ln C\right)_{,\phi}\left(1-3\frac{\delta p_c}{\delta\rho_c}\right)\delta_c\;.
\end{split}
\end{equation}
From Eq. (\ref{KG_conformal}), one immediately observes that the conformal coupling modifies the DE mass term by an effective mass term proportional to the second field derivative of the logarithm of the conformal coupling. We should further mention that the above perturbation equations can be obtained from the Newtonian gauge perturbation equations, presented in Appendix \ref{sec:App_A2}, by applying a gauge transformation \cite{Ma:1995ey,Mukhanov:1990me}. One can easily obtain the synchronous gauge perturbed coupling $\delta Q^{\text{syn}}$, by applying a gauge transformation to the corresponding perturbed coupling expression  in the conformal Newtonian gauge $\delta Q^{\text{con}}$, where these are related by $\delta Q^{\text{syn}}=\delta Q^{\text{con}} - Q^\prime k^{-2}\left(h^\prime /2+3\eta^\prime\right)$, with $Q$ being the background coupling function given by Eq. (\ref{Q}).  

\subsection*{Appendix A2: Newtonian gauge}
\renewcommand{\theequation}{A2.\arabic{equation}}
\renewcommand{\thesection}{A2}
\refstepcounter{section}
\label{sec:App_A2}
In the conformal Newtonian gauge \cite{Mukhanov:1990me}, the perturbations are characterised by the scalar potentials $\Psi$ and $\Phi$ which appear in the line element as
\begin{equation}
ds^2=a^2(\tau)\left[-\left(1+2\Psi\right)d\tau^2+\left(1-2\Phi\right)\delta_{ij}dx^i dx^j \right]\;,
\end{equation}
leading to the Newtonian gauge perturbed Einstein field equations
\begin{align}
k^2\Phi+3\mathcal{H}\left(\Phi^\prime+\mathcal{H}\Psi\right) &= -4\pi Ga^2\sum \delta\rho\;,\\
k^2\left(\Phi^\prime+\mathcal{H}\Psi\right) &= 4\pi Ga^2\sum\rho(1+w)\theta\;,\\
\Phi^{\prime\prime}+\mathcal{H}\left(\Psi^\prime+2\Phi^\prime\right)+\Psi\left(\mathcal{H}^2+2\mathcal{H}^\prime\right)+\frac{k^2}{3}\left(\Phi-\Psi\right) &= 4\pi Ga^2\sum\delta p\;,\\
k^2\left(\Phi-\Psi\right) &= 12\pi Ga^2\sum\rho(1+w)\sigma\;,
\end{align}  
where we made use of the same re--definition of the anisotropic stress as introduced in the synchronous gauge calculation. The uncoupled baryonic and radiation sectors satisfy the standard perturbed conservation equations
\begin{align}
\delta_i^\prime + 3\mathcal{H}\left(\frac{\delta p_i}{\delta\rho_i}-w_i\right)\delta_i&=-\left(1+w_i\right)\left(\theta_i-3\Phi^\prime\right)\;,
\\ \theta_i^\prime+\left[\mathcal{H}\left(1-3w_i\right)+\frac{w_i^\prime}{1+w_i}\right]\theta_i&=k^2\left[\Psi+\frac{\delta p_i}{\delta\rho_i}\frac{\delta_i}{1+w_i}\right]-k^2\sigma_i\;,
\end{align}
where $i=\{b,\;r\}$, while the perturbed evolution of a generic shear--free coupled fluid is governed by the modified perturbed continuity and Euler equations
\begin{align}
\delta_c^\prime + 3\mathcal{H}\left(\frac{\delta p_c}{\delta\rho_c}-w_c\right)\delta_c&=-\left(1+w_c\right)\left(\theta_c-3\Phi^\prime\right)+\frac{Q}{\rho_c}\phi^\prime\delta_c-\frac{Q}{\rho_c}\delta\phi^\prime-\frac{\phi^\prime}{\rho_c}\delta Q\;,
\\ \theta_c^\prime+\left[\mathcal{H}\left(1-3w_c\right)+\frac{w_c^\prime}{1+w_c}\right]\theta_c&=k^2\left[\Psi+\frac{\delta p_c}{\delta\rho_c}\frac{\delta_c}{1+w_c}\right]+\frac{Q}{\rho_c}\phi^\prime\theta_c-\frac{Q}{\rho_c\left(1+w_c\right)}k^2\delta\phi\;.
\end{align} 
The evolution of the perturbed scalar field is governed by the perturbed Klein--Gordon equation
\begin{equation}
\delta\phi^{\prime\prime}+2\mathcal{H}\delta\phi^\prime+\left(k^2+a^2V_{,\phi\phi}\right)\delta\phi=\left(\Psi^\prime+3\Phi^\prime\right)\phi^\prime-2a^2V_{,\phi}\Psi+a^2\delta Q+2a^2Q\Psi\;.
\end{equation}
In this gauge the perturbation of the coupling function $Q$, defined by Eq. (\ref{Q}), is given by \cite{vandeBruck:2012vq}
\begin{equation}\label{delta_Q}
\delta Q=-\frac{\rho_c}{a^2C+D\left(a^2\rho_c-{\phi^\prime}^2\right)}\left(\mathfrak{\widetilde{B}}_1\delta_c+\mathfrak{\widetilde{B}}_2\Phi^\prime+\mathfrak{\widetilde{B}}_3\Psi+\mathfrak{\widetilde{B}}_4\delta\phi^\prime+\mathfrak{\widetilde{B}}_5\delta\phi\right)\;,
\end{equation}
where
\begin{align}
\mathfrak{\widetilde{B}}_1=&\frac{1}{2}a^2C_{,\phi}\left(1-3\frac{\delta p_c}{\delta\rho_c}\right)-3\mathcal{H}D\left(1+\frac{\delta p_c}{\delta\rho_c}\right)\phi^\prime-a^2D\left(V_{,\phi}-Q\right)-D{\phi^\prime}^2\left(\frac{C_{,\phi}}{C}-\frac{D_{,\phi}}{2D}\right)\;,\\
\mathfrak{\widetilde{B}}_2=&3D\phi^\prime\left(1+w_c\right)\;,\\
\mathfrak{\widetilde{B}}_3=&6\mathcal{H}D\phi^\prime\left(1+w_c\right)+2D{\phi^\prime}^2\left(\frac{Q}{\rho_c}+\frac{C_{,\phi}}{C}-\frac{D_{,\phi}}{2D}\right)\;,\\
\mathfrak{\widetilde{B}}_4=&-3\mathcal{H}D\left(1+w_c\right)-2D\phi^\prime\left(\frac{Q}{\rho_c}+\frac{C_{,\phi}}{C}-\frac{D_{,\phi}}{2D}\right)\;,\\
\mathfrak{\widetilde{B}}_5=&\frac{1}{2}a^2C_{,\phi\phi}\left(1-3w_c\right)-\left(1+w_c\right)k^2D-a^2DV_{,\phi\phi}-a^2D_{,\phi}V_{,\phi}-3\mathcal{H}D_{,\phi}\left(1+w_c\right)\phi^\prime\nonumber\\
&-D{\phi^\prime}^2\left(\frac{C_{,\phi\phi}}{C}-\frac{C_{,\phi}^2}{C^2}+\frac{C_{,\phi}D_{,\phi}}{CD}-\frac{1}{2}\frac{D_{,\phi\phi}}{D}\right)+\frac{Q}{\rho_c}\left(a^2C_{,\phi}+a^2D_{,\phi}\rho_c-D_{,\phi}{\phi^\prime}^2\right)\;.
\end{align}
For the pure disformal scenario, the above perturbation equation for the coupling function simplifies to the following
\begin{equation}
\begin{split}
\delta Q^{(d)}=&\left[\left(a^2-D{\phi^\prime}^2\right)\frac{Q}{\zeta}-\frac{3\mathcal{H}D\rho_c}{\zeta}\left(w_c-\frac{\delta p_c}{\delta\rho_c}\right)\phi^\prime\right]\delta_c-3\frac{D\phi^\prime\rho_c}{\zeta}\left(1+w_c\right)\Phi^\prime\\
&-\frac{a^2\phi^\prime\rho_c}{\zeta^2}\left\{-D_{,\phi}\phi^\prime\left(1+D\rho_c\right)+2D^2\left[V_{,\phi}\phi^\prime+3\mathcal{H}\left(1+w_c\right)\rho_c\right]+6\mathcal{H}D\left(1+w_c\right)\right\}\Psi\\
&-\frac{\rho_c}{\zeta^2}\delta\phi^\prime\Big\{a^2D_{,\phi}\phi^\prime\left(1+D\rho_c\right)-D^2\left[2a^2V_{,\phi}\phi^\prime+3\mathcal{H}\left(1+w_c\right)\left(a^2\rho_c+{\phi^\prime}^2\right)\right]\Big.\\
&\Big.\;\;\;\;\;\;\;\;\;\;\;\;\;\;\;-3a^2\mathcal{H}D\left(1+w_c\right)\Big\}\\
&+\delta\phi\left\{k^2\frac{D\rho_c}{\zeta}\left(1+w_c\right)+\frac{\rho_c}{2\zeta}\left(2a^2DV_{,\phi\phi}-D_{,\phi\phi}{\phi^\prime}^2\right)\right.\\
&\left.\;\;\;\;\;\;\;\;\;\;+\frac{\rho_c}{2\zeta^2}\left[2a^2D_{,\phi}\left(a^2V_{,\phi}+3\mathcal{H}\phi^\prime\left(1+w_c\right)\right)+D_{,\phi}^2{\phi^\prime}^2\left(a^2\rho_c-{\phi^\prime}^2\right)\right]\right\}\;,
\end{split}
\end{equation}
where $\zeta=a^2+D\big(a^2\rho_c-{\phi^\prime}^2\big)$, as defined in Appendix \ref{sec:App_A1}. This expression for $\delta Q$ agrees with the equation given in Ref. \cite{Zumalacarregui:2012us} for the case of disformally coupled pressureless fluid. As in the synchronous gauge, the above equations simplify significantly in the pure conformally coupled case. Indeed, in the absence of a disformal coupling, the perturbed conservation equations for the coupled fluid reduce to 
\begin{align}
\delta_c^\prime& + 3\left(\frac{\delta p_c}{\delta\rho_c}-w_c\right)\left(\mathcal{H}+\frac{1}{2}\left(\ln C\right)_{,\phi}\phi^\prime\right)\delta_c\nonumber\\
&=-\left(1+w_c\right)\left(\theta_c-3\Phi^\prime\right)+\frac{1}{2}\left(1-3w_c\right)\left[\left(\ln C\right)_{,\phi}\delta\phi^\prime+\left(\ln C\right)_{,\phi\phi}\phi^\prime\delta\phi\right]\;,
\\ \theta_c^\prime&+ \left[\mathcal{H}\left(1-3w_c\right)+\frac{w_c^\prime}{1+w_c}+\frac{1}{2}\left(\ln C\right)_{,\phi}\left(1-3w_c\right)\phi^\prime\right]\theta_c\nonumber\\
&=k^2\left[\Psi+\frac{\delta p_c}{\delta\rho_c}\frac{\delta_c}{1+w_c}+\frac{1}{2}\left(\ln C\right)_{,\phi}\left(\frac{1-3w_c}{1+w_c}\right)\delta\phi\right]\;,
\end{align} 
while the perturbed Klein--Gordon equation simplifies as follows
\begin{equation}
\begin{split}
\delta\phi^{\prime\prime}&+2\mathcal{H}\delta\phi^\prime+\left[k^2+a^2V_{,\phi\phi}+\frac{1}{2}a^2\rho_c\left(1-3w_c\right)\left(\ln C\right)_{,\phi\phi}\right]\delta\phi\\
&=\left(\Psi^\prime+3\Phi^\prime\right)\phi^\prime-\frac{1}{2}a^2\rho_c\left(\ln C\right)_{,\phi}\left(1-3\frac{\delta p_c}{\delta\rho_c}\right)\delta_c-a^2\left[2V_{,\phi}+\left(\ln C\right)_{,\phi}\left(1-3w_c\right)\rho_c\right]\Psi\;.
\end{split}
\end{equation}

\end{subappendices}

\bibliographystyle{JHEP}
\bibliography{fullbib}

\end{document}